\documentclass{aa}

\usepackage{txfonts}

\usepackage{graphicx}	
\usepackage{amsmath}	
\usepackage{amssymb}	
\usepackage{gensymb}
\usepackage{tabulary}
\usepackage{subfigure}
\usepackage{epstopdf}
\usepackage{natbib}
\usepackage{afterpage}
\usepackage{float}
\usepackage{mathtools}
\usepackage{booktabs}
\usepackage{caption}
\usepackage{fancyvrb}
\usepackage{multirow}
\usepackage{cases}
\usepackage{tikz}
\usepackage[ruled,vlined]{algorithm2e}
\usepackage{rotating}
\usepackage{hyperref}
\usepackage[capitalise,nameinlink]{cleveref}

\usepackage{xr}		
\usetikzlibrary{matrix,shapes,arrows,positioning}



\defcitealias{cappellari2016structure}{C16}
\defcitealias{eisenstein1998hop}{EH96}
\defcitealias{stoughton2002sloan}{S02}
\defcitealias{graham2019atechnical}{Paper I}
\defcitealias{graham2019cenvironment}{Paper III}
\defcitealias{graham2019dclusters}{Paper IV}

\begin{document}


\title{SDSS-IV MaNGA: Complete census of massive slow-rotator early-type galaxy candidates and their environment in a volume-limited sample}
\titlerunning{SDSS-IV MaNGA: Complete census of massive slow-rotator candidates}

\author{Mark T. Graham$^1$\thanks{E-mail: mark.graham@physics.ox.ac.uk}
\and Michele Cappellari$^1$
\and Matthew A. Bershady$^{2,3}$
\and Niv Drory$^4$}
\authorrunning{M. T. Graham et al.}
\institute{Sub-department of Astrophysics, Department of Physics, University of Oxford, Denys Wilkinson Building, Keble Road, Oxford, OX1 3RH, UK
\and Department of Astronomy, University of Wisconsin-Madison, 475N. Charter St., Madison WI 53703, USA
\and South African Astronomical Observatory, Cape Town, South Africa
\and McDonald Observatory, The University of Texas at Austin, 1 University Station, Austin, TX 78712, USA
}

\date{Accepted XXX. Received YYY; in original form ZZZ}



\abstract
{Galaxy morphology is inextricably linked to environment. The morphology-density relation quantifies this relationship: the fraction of early type galaxies increases with increasing galaxy number density. However, optical morphology is only loosely related to the kinematic structure of galaxies, and in fact, about two thirds of elliptical (``spheroidal") galaxies are actually misclassified face-on disks, and would appear flattened if view edge on. A more robust classification is the slow/fast rotator classification which describes directly the kinematic structure. Slow and fast rotators form a bimodality in galaxy properties and are thought to follow distinct evolutionary paths, and so a kinematic morphology-density (kT-$\Sigma$) relation is more meaningful. To date, the kT-$\Sigma$ relation has only been studied in detail for a handful of nearby clusters, or across large numbers of clusters but with incomplete coverage.}
{In this work, we combine stellar kinematics obtained with the Sloan Digital Sky Survey's (SDSS) Mapping Nearby Galaxies at Apache Point Observatory (MaNGA) survey with classifications using a novel visual method to obtain the largest complete census of slow and fast rotators in nearby galaxy groups and clusters.}
{To account for incompleteness in the SDSS spectroscopic catalogue, we combine the catalogue with the photometric catalogue, which we clean using empirical criteria based only on the photometry.}
{We test our visual classification method and find our false negative rate, where a slow rotator is misclassified as a fast rotator, to be only $\sim6\%$ (0\% for $M \gtrsim 10^{11.7} \textrm{M}_{\odot}$). In contrast, our false positive rate is about 50\% implying half of all slow rotator candidates will be confirmed as fast rotators if stellar kinematics become available. Hence, our misclassification is essentially random with only a weak dependence on stellar mass, and as slow rotators are intrinsically rare, the absolute number of misclassifications will be small compared to the sample size.}
{The result is the largest complete census of massive slow rotator candidates and their environments to date.}


\maketitle



%

\section{Introduction}
Observations of galaxy clusters with IFS have revealed that there are tight relations between galaxy kinematic morphology and location within clusters \citepalias{cappellari2016structure}. This kinematic-morphology density (kT-$\Sigma$) relation \citep{cappellari2011atlas3db} is an update to the classic morphology relation of \cite{dressler1980morphology}, where instead of separating early-type galaxies (ETGs) by the traditional Hubble morphology i.e. S0s and ellipticals \citep{hubble1926extragalactic, hubble1936realm, sandage1961hubble}, ETGs are separated according to their angular momentum as revealed by their stellar kinematics obtained with IFS. ETGs can be robustly classified as fast or slow rotators by considering a proxy for the specific stellar angular momentum $\lambda_{R_e}$ within the half-light radius $R_e$ and the galaxy apparent flattening $\epsilon$ \citep{emsellem2007sauron, emsellem2011atlas3d, cappellari2016structure}. The two classes of ETG are thought to be the end result of two distinct formation channels \citepalias{cappellari2016structure}, and observational evidence has shown that they form a bimodality in galaxy properties, rather than a smooth continuum \citep{graham2018angular}.\par
Previous studies of the kT-$\Sigma$ relation fall into two categories:
\begin{enumerate}
\item Three, very nearby and well-studied clusters were mapped in detail with IFS producing a complete census of slow rotators (SRs). These three clusters were the Fornax \citep{scott2014distribution} and Virgo \citep{cappellari2011atlas3db} clusters as well as the central part of the Coma \citep{houghton2013densest} cluster. Schematics for these clusters are presented in fig. 26 of \citetalias{cappellari2016structure} where the genuine dry merger relics, i.e. core SRs above a critical mass of $M_{\rm{crit}} \geq 2 \times 10^{11} \textrm{ M}_{\odot}$ \citep{cappellari2013effect}, are highlighted. 
\item Large-scale IFS surveys have mapped large numbers of clusters and groups, but with only very sparse coverage. This ``incomplete'' approach was started by \cite{houghton2013densest, d2013fast} and \cite{scott2014distribution} and was later extended by \cite{brough2017kinematic} with data from SAMI \citep{croom2012sydney} and by \cite{greene2017kinematic} with data from the Mapping Nearby Galaxies at Apache Point Observatory (MaNGA) survey \citep{bundy2015overview} (see \citealp{smee2013multi} for details about the spectrographs, \citealp{drory2015manga} for a complete description of the IFUs, \citealp{law2015observing} for details about the observing strategy and \citealp{yan2016spectrophotometric} for details about the flux calibration). Both surveys encompass a broad range of environments from clusters and groups to the field and provide up to two orders of magnitude more galaxies than the dedicated cluster observations.
\end{enumerate}
Here, we try for the first time to combine the strength of both approaches by mapping a large number of groups and clusters while making a complete census of the massive SRs in those clusters. Such a census would allow us to explore the generality of the clear result seen in fig. 26 of \citetalias{cappellari2016structure} where the massive SRs are seen to lie at the centres of clusters, or in some cases at a secondary peak within clusters.\par
We base our study around a sample of about 4600 galaxies observed by the MaNGA survey. The survey targets galaxies in the redshift range $0.01\leq z \leq 0.15$ and has a flat selection in stellar mass, making it the best sample currently available for studying massive SRs, with about \textit{1200} ETGs above $M_{\rm{crit}}$. The nearest competitor, the MASSIVE survey, only has about 100 galaxies above $M\sim10^{11.5} \textrm{ M}_{\odot}$, while MaNGA has nearly \textit{900} ETGs in the same mass range. SAMI only has a handful of galaxies in the same mass range (e.g. \citealp{brough2017kinematic}). The targets for the MaNGA survey are taken from the NASA-Sloan Atlas (NSA; \citealp{blanton2011improved}) and so we could rely on the NSA to finding neighbouring galaxies of MaNGA galaxies. However, the NSA is not complete (\autoref{sec:photometry}) and so we decide to supplement it with the SDSS photometric catalogue (\citealp{stoughton2002sloan}; hereafter S02). However, the photometric catalogue does not present a comparable level of accuracy to the NSA in the sense that the proportion of objects which are not galaxies is higher in the photometric catalogue compared to the NSA. Hence, in \autoref{sec:photometry} we ``clean'' up the catalogue using some empirical criteria for selecting clean photometry that cannot otherwise be selected by using the flags generated by the SDSS imaging pipeline \citepalias{stoughton2002sloan}.\par
As we require stellar mass estimates for all galaxies in our combined catalogue, we use dynamical mass estimates for MaNGA galaxies to estimate the stellar mass from the absolute luminosity in the SDSS $r$-band. (\autoref{sec:photometry_mass}). In \autoref{sec:group_finder}, we describe our method for finding galaxy groups that contain MaNGA galaxies using the group finder algorithm, \texttt{TD-ENCLOSER}, which is described in detail in \cite{graham2019atechnical} (Paper I). Finally, we identify candidate massive slow rotators in a redshift-limited sample and quantify our performance by performing the same classification on MaNGA galaxies where we know the true fast/slow classification (\autoref{sec:srcandidates}).  In two follow-up papers, we use the catalogue presented here to conduct a large study of galaxy angular momentum and environment (\citealp{graham2019cenvironment}, Paper III) before focussing on a few specific examples (\citealp{graham2019dclusters}, Paper IV).
\section{Cleaning up the SDSS photometric catalogue}
\label{sec:photometry}
In the case of MaNGA, there are two potential sources of incompleteness which may hamper any attempt to undertake a study of the kT-$\Sigma$ relation in a way that emulates the detailed studies of nearby clusters. The first is a direct consequence of the limited scope of the SDSS spectroscopic sample. Spectroscopy provides an accurate redshift estimate, and hence the location of galaxies with spectroscopic observations is well-defined in three dimensions. The disadvantage of spectroscopic observations is that they are time consuming and are limited to the brightest galaxies at higher redshifts. The SDSS spectroscopic sample is at least 95\% complete above an apparent magnitude of 17.77 mag in the $r$-band. This results in a difference of a factor of 250 in the minimum intrinsic luminosity that is possible to observe between $z=0.01$ and $z=0.15$ for example. This change in the sampling of the luminosity function with redshift means that the low-mass neighbours of galaxies at the higher redshifts within the SDSS would not be observed with spectroscopy. Furthermore, the finite size of the fibres means that it is not possible to obtain spectroscopic redshifts for two galaxies if the distance between them is less than 55 arcsec on the sky (a case which is referred to as a fibre collision, see \citealp{blanton2003sloan}). This effect is particularly enhanced in dense cluster regions (and for close pairs) and so it presents a serious problem for detailed studies of cluster environment with spectroscopy. In total, about 6\% of all objects that are identified by the bespoke SDSS imaging pipeline \citepalias{stoughton2002sloan} to be galaxies \textit{and} are brighter than $m_r=17.77$ mag are not observed with spectroscopy \citep{strauss2002sloan}, although this fraction was recently estimated by \cite{lazo2018sdss} to be 7.2\%.\par
One possible way to overcome this incompleteness would be to apply a weighting to the group luminosity or richness (multiplicity) as a function of redshift as was done by \cite{tempel2012groups}. While it would be possible to weight other parameters such as the projected surface density $\Sigma_n \equiv n/(\pi D_n^2)$, which considers the proximity $D_n$ of the nearest $n$th neighbour to a galaxy, it would not be possible to recover the intrinsic environment of these galaxies at $z\sim0.1$. To gain the most information about how a galaxy is affected by its environment, the optimal strategy is to actually \textit{observe} its environment, rather than look at trends as a function of a suitable parameter such as $\Sigma_n$.\par
Here we take an entirely different approach by combining the SDSS spectroscopic sample with the photometric catalogue obtained with imaging \citep{gunn20062, york2000sloan}. For our spectroscopic sample, we use the final version of the NASA Sloan Atlas (\texttt{v1\_0\_1}), released as part of DR13 \citep{albareti2016thirteenth}. The NSA is an improvement on the standard SDSS pipeline with better sky subtraction for larger galaxies and a better treatment of deblending \citep{blanton2011improved}. It contains spectroscopic and photometric data for 636,394 galaxies below $z \approx 0.15$. The SDSS DR12 photometric catalogue \citep{alam2015twelve} is 95\% complete for point sources at magnitudes brighter than $m_r=22.2$ \citepalias{stoughton2002sloan} and as a result, it is two orders of magnitude larger than the spectroscopic sample. By combining the NSA with the photometric catalogue in a robust way, we can construct a large galaxy catalogue that doesn't suffer from the same selection effects and biases that the NSA does.\par

\subsection{Summary of the SDSS imaging pipeline}
\label{sec:imaging}
The SDSS imaging pipeline is the powerhouse of the photometric catalogue and performs object detection, deblending and flagging \citepalias{stoughton2002sloan}. Objects are initially detected above 200$\sigma$ where $\sigma$ is the local sky level after the global sky background has been subtracted. Afterwards, the image is smoothed at the level of the PSF ($1\arcsec.2-2\arcsec.3$ at full-width at half maximum) and objects are then detected above 5$\sigma$. After detection, the objects are ``grown'' in increments equal to the size of the PSF. Objects are deblended if multiple individual peaks, including the original peak, are detected across the five SDSS $ugriz$ bands. These multiple peaks are referred to as ``children'' of the ``parent'' (original), and the parent is considered obsolete from then on.\par
The pipeline uses a simple criterion to separate stars (which are point-like) and galaxies (which are extended objects). Each object has a \texttt{psfMag} quantity, which is the flux within a Gaussian model of the PSF centred on the object (see sec. 4.4.5.2 of \citetalias{stoughton2002sloan}), and a \texttt{cModelmag} quantity, which is the magnitude from a fit produced from a linear combination of the de Vaucouleurs \citep{de1948extragalactic} and exponential \cite{sersic1968atlas} fits. If \texttt{psfMag} is large enough compared to \texttt{cModelMag}, then the object is classified as a star (see sec. 4.4.6.1 of \citetalias{stoughton2002sloan}).\par
The pipeline has been improved since Data Release 1 \citep{abazajian2003first} to better deal with unusual cases (such as foreground stars). However, \citetalias{stoughton2002sloan} acknowledged that the automatic star-galaxy separation has a 95\% accuracy rate at $m_r=21$ mag and, as there are $2\times 10^8$ objects in the SDSS DR12, there are likely to be objects which on closer inspection appear to be misclassified. In our assessment, we find many objects that have been classified by the imaging pipeline to be galaxies with clean photometry, but actually appear to the naked eye to have a different appearance. In this section, we describe two such cases (\autoref{sec:faint} and \autoref{sec:bright}). At first, we use colour, magnitude and half-light radius to classify objects. We also include the error in photometric redshift as this error correlates with the error in flux (\autoref{sec:photoz}). Finally, we show that our criteria are independent of the SDSS flags assigned by the pipeline. Before we present our findings, we describe how we obtain a clean selection of objects from the photometric catalogue.\par

\subsection{Initial selection from the database}
\label{sec:clean}
We obtain a clean sample of objects that have been classified as galaxies from the SDSS DR12\footnote{The latest data release is DR15 \citep{aguado2018fiftheenth} which contains identical imaging to DR12.} photometric catalogue \citep{alam2015twelve} using the following SQL query on CasJobs:

\begin{Verbatim}[tabsize=4]
SELECT g.ra, g.dec, g.petroR50_r,
	gcpetroMag_u, g.petroMagErr_u,
	g.petroMag_g, g.petroMagErr_g,
	g.petroMag_r, g.petroMagErr_r,
	g.petroMag_i, g.petroMagErr_i,
	g.petroMag_z, g.petroMagErr_z,
	g.expAB_r,
	pz.z, pz.zErr, 
	pz.photoErrorClass, pz.nnCount 
FROM GalaxyTag AS g
JOIN Photoz AS pz
ON g.ObjID = pz.ObjID
WHERE
	g.type = 3                        # (1)
	AND g.clean = 1                   # (2)
	AND (g.calibStatus_r & 1) != 0    # (3)
\end{Verbatim}

For each object, we select the RA and Dec coordinates, the Petrosian half-light radius in the $r$-band \texttt{petroR50\_r} \citep{petrosian1976galaxies}, the apparent magnitude and errors from the Petrosian fit in the $ugriz$ bands (\texttt{petroMag\_u}, \texttt{petroMagErr\_u} etc.), the axial radio of an exponential fit to the $r$-band (\texttt{expAB\_r}), the redshift and redshift error (\texttt{z}, \texttt{zErr}) as well as two quantities related to the redshift (\texttt{photoErrorClass} and \texttt{nnCount}; see \autoref{sec:photoz}).\par
Objects can be detected more than once if they lie in regions where two or more scans overlap. To avoid selecting duplicates, we only select objects in the \texttt{GalaxyTag} database, which contains objects which have been classified as ``primary''. Next, we select objects with \texttt{g.type = 3} ((1) in the SQL query) which identifies objects that have been classified as galaxies by the automatic pipeline (sec \autoref{sec:imaging}). The pipeline generates flags for each object in the catalogue. To aid the user in quickly selecting objects with clean photometry, a shortcut is available that is designed to select only objects which have reliable photometry. The shortcut, which can be applied using (2) and (3) in the SQL query, is described on the SDSS website\footnote{\url{http://www.sdss.org/dr12/algorithms/photo_flags_recommend/}}, but here we briefly summarise its function. Each selection below is made in the $r$-band which has the best signal-to-noise of the five SDSS bands.\par
The function of \texttt{g.clean = 1} is to:
\begin{itemize}
\item only select objects which do not suffer from deblending problems in the $r$-band.
\item check if there have been substantial interpolation issues in the $r$-band. If so, the object has too few good pixels to estimate the photometric errors from and so the errors may be underestimated \citepalias{stoughton2002sloan}.
\item ensures that objects are detected in the first pass where the image is only smoothed to the level of the PSF. (Many objects that are first detected after 2$\times$2 or 4$\times$4 pixel binning are not astrophysical objects or scattered light from stars.)
\item check that objects are not saturated as the photometry can be inaccurate.
\item check if an object has a valid radial profile, otherwise the photometric quantities are likely to be unreliable.
\end{itemize}
An extra criterion is that \texttt{calibStatus\_r} flag is set to true, indicating that the calibration in the $r$-band is successful. The flags are described in detail in the Early Data Release paper \citepalias{stoughton2002sloan}. By applying the recommended shortcut, the objects obtained are not affected by any of these issues. Of the objects that satisfy these criteria, we find two populations, one faint and one bright, which appear erroneous. We suggest simple empirical criteria that are defined using the photometry and can be used to remove these objects. To obtain a test catalogue, we run the above query within a region of sky defined by $200\degree \leq \textrm{RA} \leq 210\degree$ and $30\degree \leq \textrm{Dec} \leq 40\degree$.\par

\subsection{Removing spurious faint objects}
\label{sec:faint}
\begin{figure*}
\centering
\includegraphics[page=1, width=0.95\textwidth]{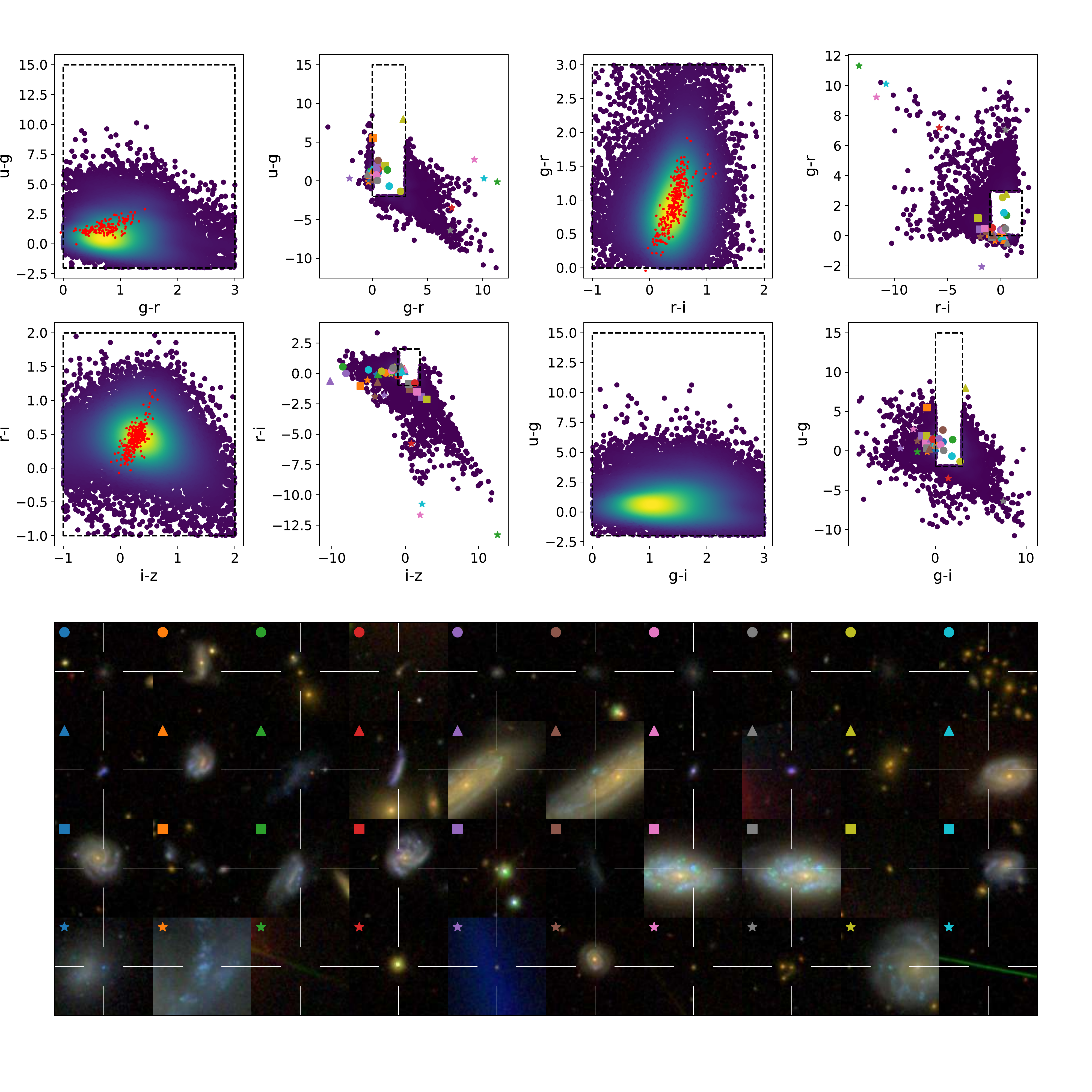}
\caption[Colour-colour plots for objects in the SDSS photometric catalogue]{\textbf{Upper:} Colours of objects in the photometric catalogue with $R_{\rm{r,Petro}} \leq 10\arcsec$. In the first two panels (i.e. top left), we plot $u-g$ versus $g-r$ for 20,000 randomly selected objects of all magnitudes within a 10x10 degree patch of the sky. Of these two panels, the first contains all objects that fall within the bounding box indicated by the dashed lines, and the second contains all objects that fall outside this box. Each point is coloured by the density estimated using a KDE. In both panels, the colour scale is identical. The remaining three sets of panels plot the same for different colour combinations: $g-r$ v $r-i$, $r-i$ v $i-z$ and $u-g$ v $g-i$. Note that the objects that appear inside/outside each box are not necessarily the same in all panels. In some of the panels, we plot colours for about 300 galaxies from  \protect\cite{newberg1999color} (red crosses). 
\textbf{Lower:} Optical images from the SDSS for 40 randomly chosen objects that fail to satisfy all the colour-colour criteria, and hence lie outside \textit{at least} one of the boxes. Each object has a coloured marker associated with it which can be used to match the object to the four wide-view panels in the upper section. Cross-hairs are included to indicate the centre to guide the eye.}
\label{fig:color-color}
\end{figure*}

In the photometric catalogue, we find many objects that have unusual colours. We also find many objects which have been incorrectly fitted with an unrealistically large radius which produces an artificial overestimation of the total magnitude and consequently are recorded to be much brighter than they appear. We suspect they are likely to be reddened galaxies at high redshift. To exclude these and other unusual objects, we examine the distribution of objects on a colour-colour plane using various combinations of the five SDSS $ugriz$ bands. We use the four traditional SDSS colours, $u-g$, $g-r$, $r-z$ and $i-z$, as well as $g-i$ to provide a fifth colour. We find that the vast majority of galaxies lie close to a peak in the five-dimensional colour space (see the top panels in \autoref{fig:color-color}). The peaks of the colour distributions are roughly between 0-2 suggesting that the fluxes in both bands are comparable. The location of the peaks agrees broadly with that found by \cite{hansson2012photometry} who studied a smaller, brighter sample of SDSS galaxies. However, we find a significant population of outliers which can extend out to extreme parts of the parameter space. Many of these objects are detected in one band but are undetected in another band. We find that many of these objects are in fact artifacts (see the images in \autoref{fig:color-color}).\par

We use the distribution of objects in the five-parameter colour space to define empirical criteria for selecting clean objects:

\begin{align}
-2 &\leq u-g \leq 15 \nonumber \\
0 &\leq g-r \leq 3 \nonumber \\
-1 &\leq r-i \leq 2  \nonumber \\
-1 &\leq i-z \leq 2 \nonumber  \\
0 &\leq g-i \leq 3 
\label{eq:colour_criteria}
\end{align}
In \autoref{fig:color-color}, we plot various combinations of the five colour criteria for all galaxies that satisfy the other selection criteria in this section. Rather than choosing to keep a specific percentile of objects, we simply use the density peaks to define our criteria. We ensure that each boundary is far enough from the peak so as only to clip outliers. We only keep photometric objects that satisfy \textit{all} of these colour criteria. For $15 \leq m_r<18$, at least $\sim99\%$ of objects satisfy these criteria, while for $14 \leq m_r<15$ and $18 \leq m_r<19$, this fraction is $\sim97\%$. For bins $13 \leq m_r<14$ and $19 \leq m_r<20$, $\sim96\%$ and $\sim93\%$ survive the colour cut respectively. The errors in $u$-band are greater than the other bands because the flux calibration is accurate to 2\% in this band compared to the other bands \citep{padmanabhan2008sloan}, and so we have a relaxed criteria for this band. These empirical criteria exclude unusual objects with only a minimal chance of removing genuine galaxies. To illustrate this in \autoref{fig:color-color}, we overplot galaxies from \cite{newberg1999color} who measured the photometric properties of different celestial objects using the SDSS photometry (see \citealp{fan1999color} for simulations). For each combination of colours, the peak is strongly centred cleanly inside the bounding box. There is structure outside the bounding boxes which correspond to different types of object (see \citealp{hansson2012photometry} for some examples).\par
We randomly select 40 objects which fail to satisfy at least one of the criteria in \cref{eq:colour_criteria} and show images in \autoref{fig:color-color}. Overall, we find that some of the objects are faint, while others are duplicates of objects or cases where it is ambiguous where the brightest part of the object is. In a small number of cases, objects that have genuine close neighbours in a cluster environment ($\lesssim 5$ arcsec) are excluded on the basis of their $i-z$ colour. Our colour criteria are chosen purely to exclude as many outliers as possible while keeping as many genuine galaxies as possible. As the selection is empirical, we acknowledge that it is not perfect. However, as the vast majority of objects are not close to the boundaries, the chances of a normal galaxy being unfairly excluded are very small.\par
As an aside, the $u-r$ colour is used to separate galaxies into a red and blue sequence \citep{strateva2001color,baldry2004bimodal}. A colour of $u-r\approx2$ is used as a divider, and in the $(g-r,u-g)$ plane, there are two peaks either side of the diagonal line. However, in the first panel in \autoref{fig:color-color} comparing $g-r$ with $u-g$, there is only one peak visible. The dividing line varies as a function of absolute magnitude \citep{baldry2004bimodal} and moves towards bluer colours with decreasing flux. As the number counts of galaxies increases with decreasing flux, faint galaxies vastly outnumber bright galaxies in \autoref{fig:color-color}. As faint galaxies tend to be in the blue sequence, only this peak is visible. To check, we replot $g-r$ vs $u-g$ for $m_r < 18$ (not shown) and find the peak corresponding to the red sequence.\par
We note that \cite{lazo2018sdss} have also attempted to clean the SDSS photometric catalogue using colour criteria. They adopted a visual classification approach not dissimilar to the one used here to eliminate artifacts and spurious objects. They find a slightly higher incompleteness of 7.2\% than the 6\% quoted by \citetalias{stoughton2002sloan}.

\begin{figure*}
\centering
\includegraphics[width=0.95\textwidth]{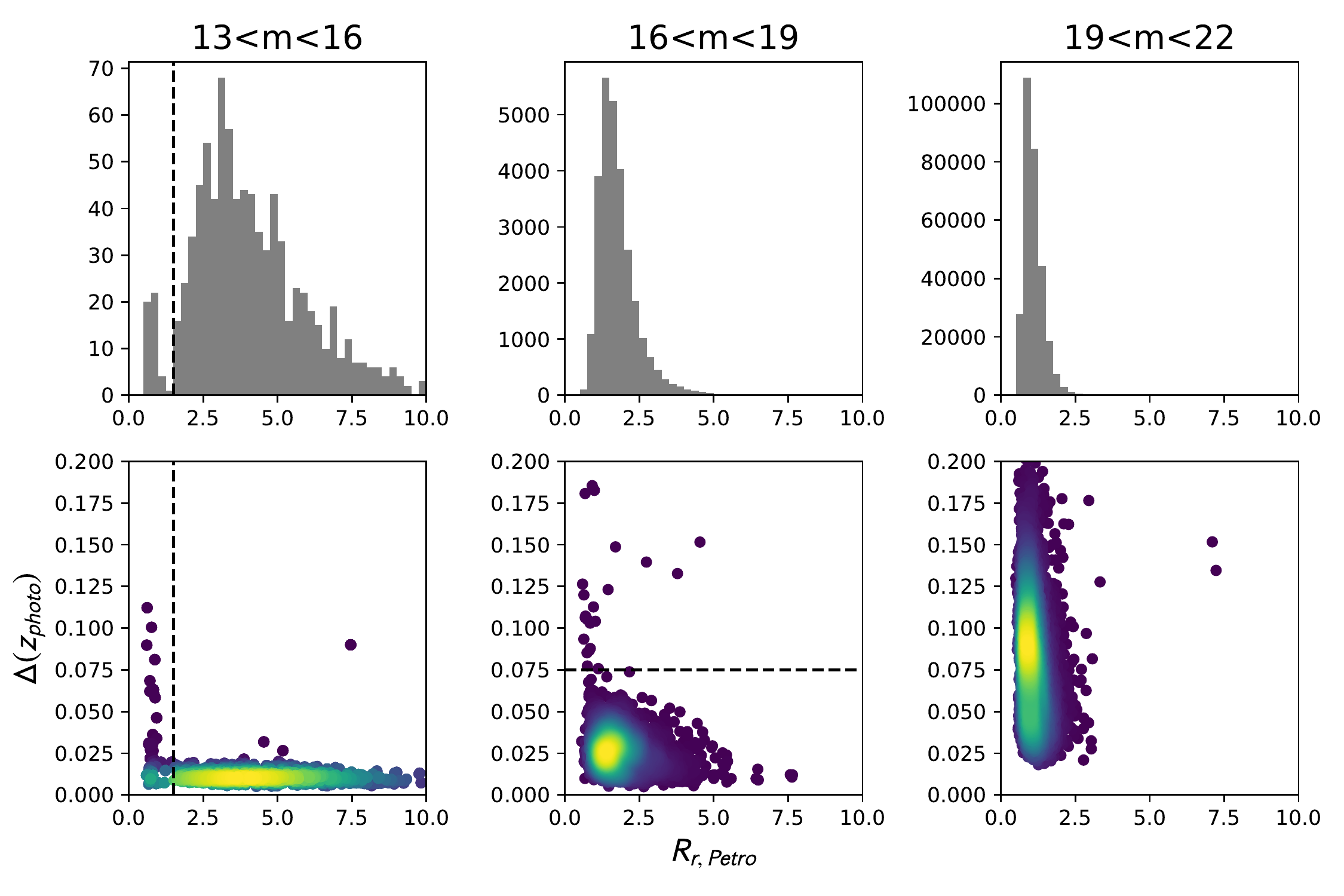}
\caption[Distribution of $R_{\rm{r,Petro}}$ and $z_{\rm{err}}$ as a function of $m_r$]{\textbf{Top:} The distribution of the Petrosian $r$-band half-light radius $R_{\rm{r,Petro}}$ for three magnitude bins. The vertical line in the first panel indicates a cut in radius. \textbf{Bottom:} The error in photometric redshift versus $R_{\rm{r,Petro}}$ for each magnitude bin. In the left and middle panel, we show the region where we exclude galaxies ($R_{\rm{r,Petro}}<1.5$ arcsec and $z_{\rm{err}} > 0.075$ respectively). The density of points is indicated by the colourmap.}
\label{fig:re_cuts}
\end{figure*}

\subsection{Removing spurious bright objects}
\label{sec:bright}

\begin{figure*}
\centering
\includegraphics[width=0.95\textwidth]{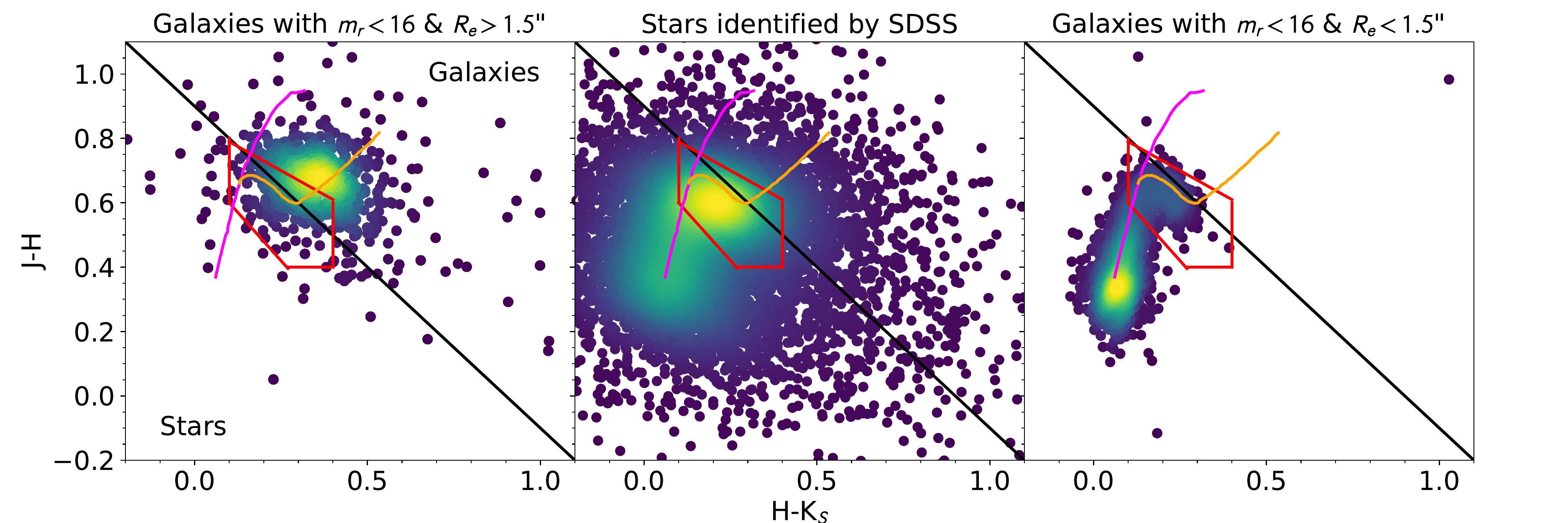}
\caption[$J-H$ versus $H-K_S$ infrared colours from the 2MASS PSC for objects that satisfy our colour cuts.]{\textbf{$J-H$ versus $H-K_S$ infrared colours from the 2MASS PSC for objects that satisfy our colour cuts.} \textbf{Left:} $J-H$ versus $H-K_S$ for objects larger than 1.5 arcsec and brighter than $m_r=16$. Each point is coloured by the density estimated using a KDE. The solid black line indicates the dividing line between galaxies and stars ($J-K_S=0.9$). The magenta and orange tracks indicate where K0 and M0 dwarf stars appear and are taken from \protect\cite{bessell1988colours}. The region bound by the red lines is the ``red dwarf box'' from \protect\cite{lepine2011catalog}. The $J-K_S$ colours for the majority of these objects are consistent with the infrared colours of galaxies. \textbf{Middle:} The same for objects identified as stars by the SDSS imaging pipeline. The $J-K_S$ colours of these objects are consistent with the infrared colours of stars. The dominant peak corresponds to sub-dwarfs. \textbf{Right:} The same for objects identified as galaxies by the SDSS imaging pipeline. The infrared colours of these objects are almost exclusively consistent with the infrared colours of stars. The sharp peak corresponds to K0 dwarf stars.}
\label{fig:2MASS_colors}
\end{figure*}

\begin{figure*}
\centering
\includegraphics[page=2, width=0.95\textwidth]{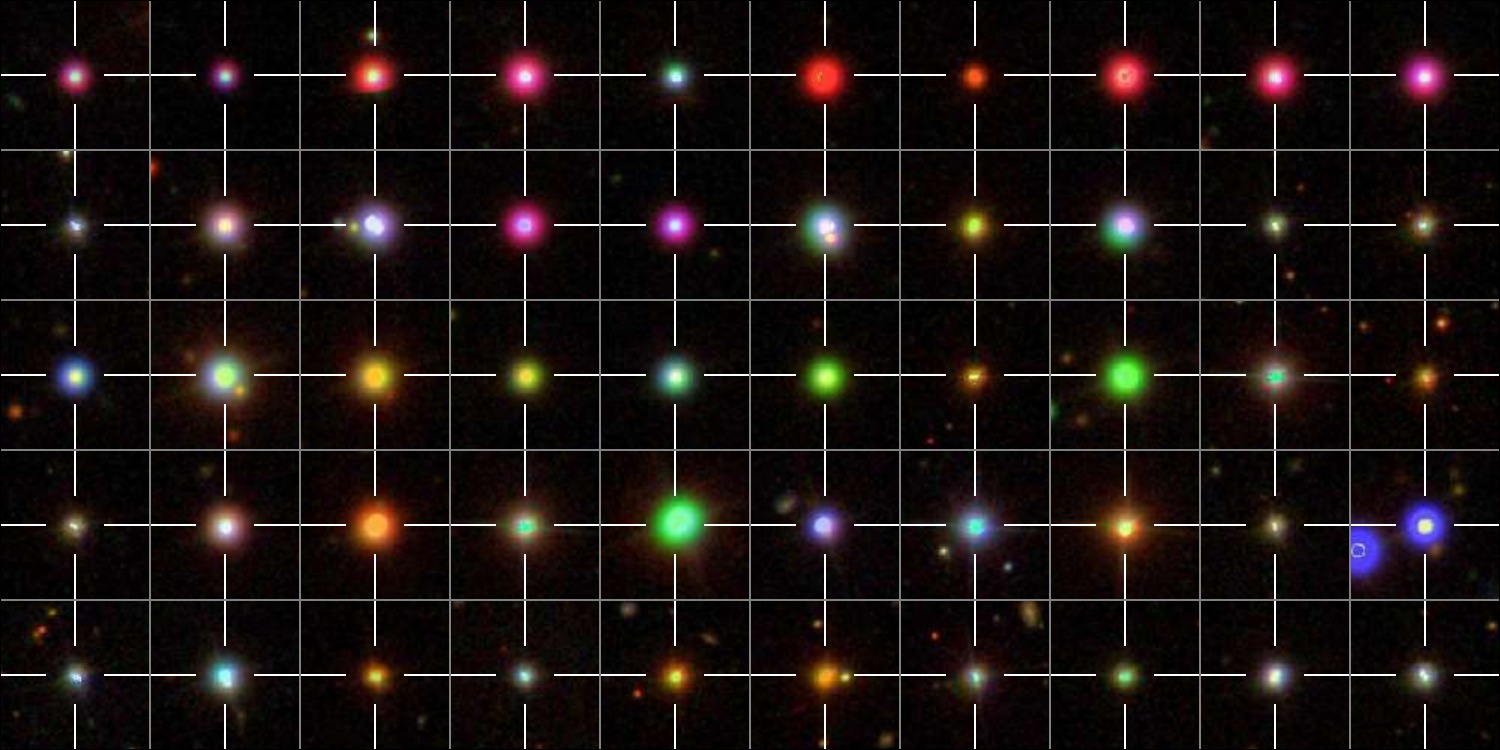}
\caption[Images of objects in SDSS and the 2MASS PSC with $R_{\rm{r,Petro}} < 1.5$ arcsec and $m_r\leq16$]{Optical images from the SDSS of objects which are smaller than $R_{\rm{r,Petro}}=1.5\arcsec$, brighter than $m_r=16$ and are found in the 2MASS PSC. All of these objects lie in the bottom-right panel of \autoref{fig:2MASS_colors} and their infrared colours are consistent with stars.}
\label{fig:2MASS_stars}
\end{figure*}

In the top-left panel of \autoref{fig:re_cuts}, we plot the distribution for the Petrosian radius in the $r$-band, $R_{\rm{r,Petro}}$, for bright objects that have been classified by the pipeline to be galaxies and satisfy all the clean photometry cuts \textit{and} our empirical colour cuts. We find two peaks, one major and one minor, separated at approximately $R_{\rm{r,Petro}}=1.5$ arcsec. In \autoref{fig:2MASS_stars}, we show images for a random selection of the objects in the minor peak: These objects resemble stars in their appearance. We test the hypothesis that the two peaks correspond to galaxies and stars by obtaining magnitudes in the $J$, $H$ and $K_s$  infrared bands from the Two Micron All-Sky Survey (2MASS) Point Source Catalogue (PSC; \citealp{skrutskie2006two}). It is known that different stellar populations occupy different regions on the $(J-H,H-K_S)$ plane (see Fig. 1 of \citealp{koornneef1983colours} and Fig. A3 of \citealp{bessell1988colours}). Furthermore, $J-K_S = 0.9$ has been found to be a good discriminant for stars and galaxies, as galaxies are generally redder than this value in the infrared, although bulgeless spirals can have $J-K_S\sim0.8$ \citep{jarrett20032mass}. Interstellar reddening can make stars appear to have reddened $J-K_S$ colours, but as we are not observing through the Galactic plane, we do not expect this effect to be strong.\par

To investigate the reliability of this simple criterion, we compare the infrared colours $J-H$ and $H-K_S$ for objects that have been classified as galaxies, are brighter than $m_r=16$ mag, and have $R_{\rm{r,Petro}} \geq 1.5\arcsec$ (left panel of \autoref{fig:2MASS_colors}). We find that the peak in the distribution is placed cleanly above the demarcation line with only a small fraction lying below the line. We repeat the exercise for objects which have been classified as stars by the SDSS pipeline and find that the overall majority have $J-K_S < 0.9$ (middle panel of \autoref{fig:2MASS_colors}). Finally, we take the objects that have been classified as galaxies and are brighter than $m_r=16$ mag, but resemble stars ($R_{\rm{r,Petro}}<1.5$ arcsec) and find that they in fact peak at $J-K_S\approx0.3-0.4$, with very few appearing on the side pertaining to galaxies (right panel of \autoref{fig:2MASS_colors}). Their location of the peak matches the track belonging to the K0 dwarf stars on the diagram (magenta line; \citealp{bessell1988colours}) and the distribution also follows that occupied by M0 stars (orange line). These are predominantly red stars (see \autoref{fig:2MASS_stars}) which may explain why they were missed by the pipeline. The peak in the distribution of stars which the pipeline \textit{did} identify correctly lies within the ``red dwarf box'' of \cite{lepine2011catalog} (red box).\par


We conclude that these objects are indeed stars and should be removed from any photometric galaxy catalogue. Hence, for $m_r<16$, we require that all objects have $R_{\rm{r,Petro}}\geq1.5\arcsec$. We cannot repeat this exercise for $m_r \gtrsim 16$ for two main reasons. Firstly, galaxies with fainter apparent magnitudes are likely to be at further distances and hence will have smaller angular size (e.g. the right panel of \autoref{fig:re_cuts}). Secondly, the magnitude limit of 2MASS prevents us from getting measurements of the infrared colours for faint objects. However, we do not think that many nearby bright stars will be faint enough to lie in this magnitude range, and hence we are confident that we are not missing many other stars.\par
In the next subsection, we describe the photometric redshift estimates which we use as part of our criteria.

\subsection{Photometric redshift estimates}
\label{sec:photoz}
Photometric redshifts were estimated for all $\sim 2 \times 10^8$ objects in the photometric catalogue by \cite{beck2017redshift}. They used an empirical local linear regression technique along with a supervised machine learning algorithm to find patterns in a filtered training set of objects with spectroscopic redshifts. Each photometric object was assigned to a class which allows a predefined maximum for the error in the $r$-band magnitude as well as in the $g-r$, $r-i$ and $i-z$ colours (see their Table 2). Photometric objects were matched to spectroscopic galaxies in a five dimensional colour-magnitude parameter space using the photometric colour and magnitudes. Galaxies which are close in the parameter space are assumed to be close in redshift. The redshift estimates that are obtained for the photometric objects follow a one-to-one relationship with the redshifts of the training set up until $z \approx 0.6$ (four times our maximum redshift) beyond which the accuracy worsens rapidly (see their fig. 3). We follow \cite{aragon2015redshift} and only keep redshifts with $\texttt{nnCount} \geq 95$. A lower value of \texttt{nnCount} indicates that the galaxy does not overlap well with the training set in colour-magnitude space\footnote{\label{Photoz}\url{https://www.sdss.org/dr12/algorithms/photo-z/}}. The reliability of a given galaxy's redshift estimate can also be assessed by consulting the \texttt{photoErrorClass} parameter. A value of 1 is the best, while a value of 7 is the worst. This parameter is positive if the galaxy is within the bounding box of its 100 nearest neighbours, and negative if otherwise (because it is an outlier). It is recommended only to keep estimates with class 1 (see \autoref{Photoz}). However we also include classes -1, 2 and 3 (see \autoref{Photoz}) as the bias for these classes is close enough to class 1 (see Table 3 of \citealp{beck2017redshift}).\par
\begin{figure*}
\centering
\includegraphics[width=0.95\textwidth]{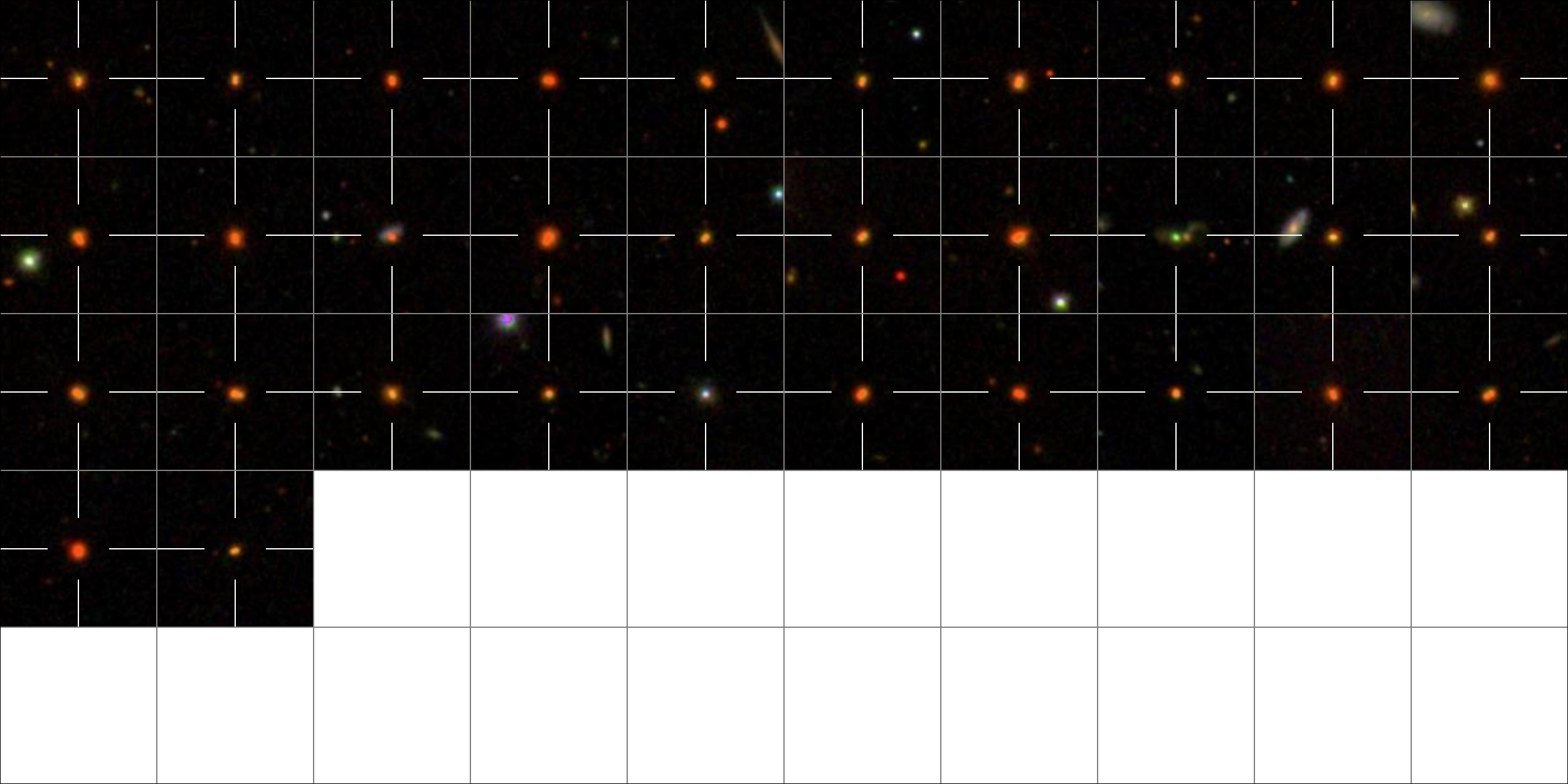}
\caption[Optical images from the SDSS for all objects with $16 \leq m_r <19$, $\Delta z_{\rm{phot}} > 0.075$ and $R_{\rm{r,Petro}} < 1.5$ arcsec]{Optical images from the SDSS for all objects with $16 \leq m_r <19$, $\Delta z_{\rm{phot}} > 0.075$ and $R_{\rm{r,Petro}} < 1.5$ arcsec. Almost all of them appear to be stars with red optical colours.}
\label{fig:high_zerr}
\end{figure*}

We check the next magnitude bin $16 < m_r \leq 19$ and find that there are some outliers which have an error in the photometric redshift $\Delta(z_{\rm{phot}}) \geq 0.075$ (middle panel of \autoref{fig:re_cuts}). We find that these are red and appear to be stars (see \autoref{fig:high_zerr}), so we also exclude this small number of objects.

\begin{figure}
\centering
\includegraphics[page=1, width=0.45\textwidth]{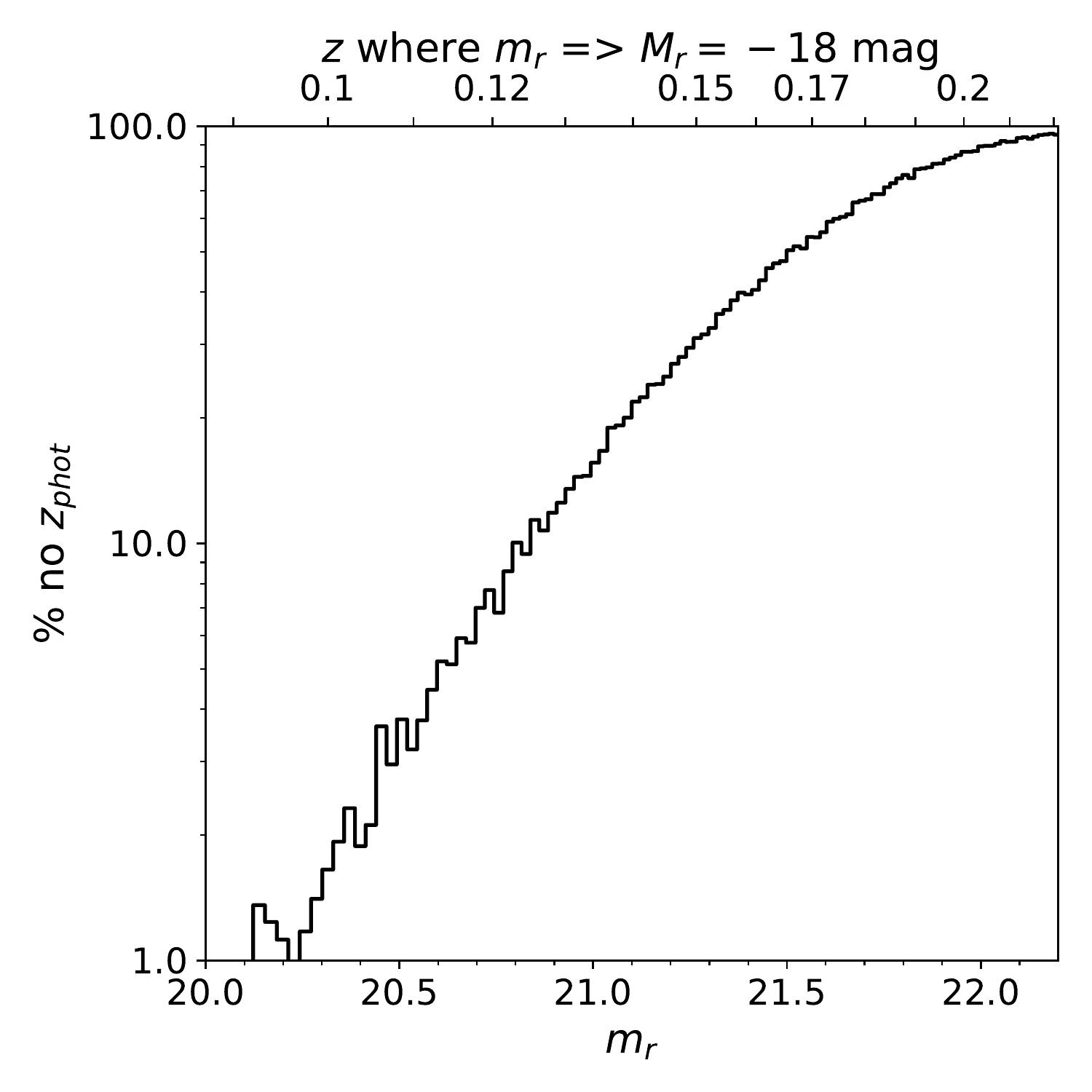}
\caption[The fraction of galaxies that do not satisfy two photometric redshift criteria]{The fraction of galaxies that do not satisfy the two photometric redshift criteria as a function of apparent $r$-band magnitude (bottom axis). On the top axis, we give the redshift where the apparent magnitude on the bottom axis corresponds to an absolute magnitude $M_r=-18$ mag (see \autoref{sec:photometry_mass}).}
\label{fig:frac_z_zphot}
\end{figure}

In \autoref{fig:frac_z_zphot}, we plot the fraction of objects which do not satisfy these two criteria as a function of apparent $r$-band magnitude. We find that at magnitudes brighter than $m_r \approx 20$, this fraction is less than 1\%. (The fraction increases towards the bright end as well but for different reasons.) This fraction increases smoothly with decreasing flux until it reaches almost 100\% at the 95\% completeness limit for point sources ($m_r=22.2$ mag). We also give the redshift where the apparent magnitude $m_r$ corresponds to an absolute luminosity $M_r = -18$ mag (see \autoref{sec:photometry_mass}).\par

\subsection{Comparison with SDSS flags}
\label{sec:sdss_flags}
Apart from selecting the type corresponding to galaxies and applying the recommended shortcut for selecting clean photometry, we have not used any other flags from the SDSS pipeline. It is important to check that we are not missing any crucial flags, as well as make sure that our cuts in photometry cannot also be obtained by simply using one or more of the SDSS flags (and hence are independent of the SDSS flags).\par
\begin{figure*}
\centering
\includegraphics[page=1, width=0.95\textwidth]{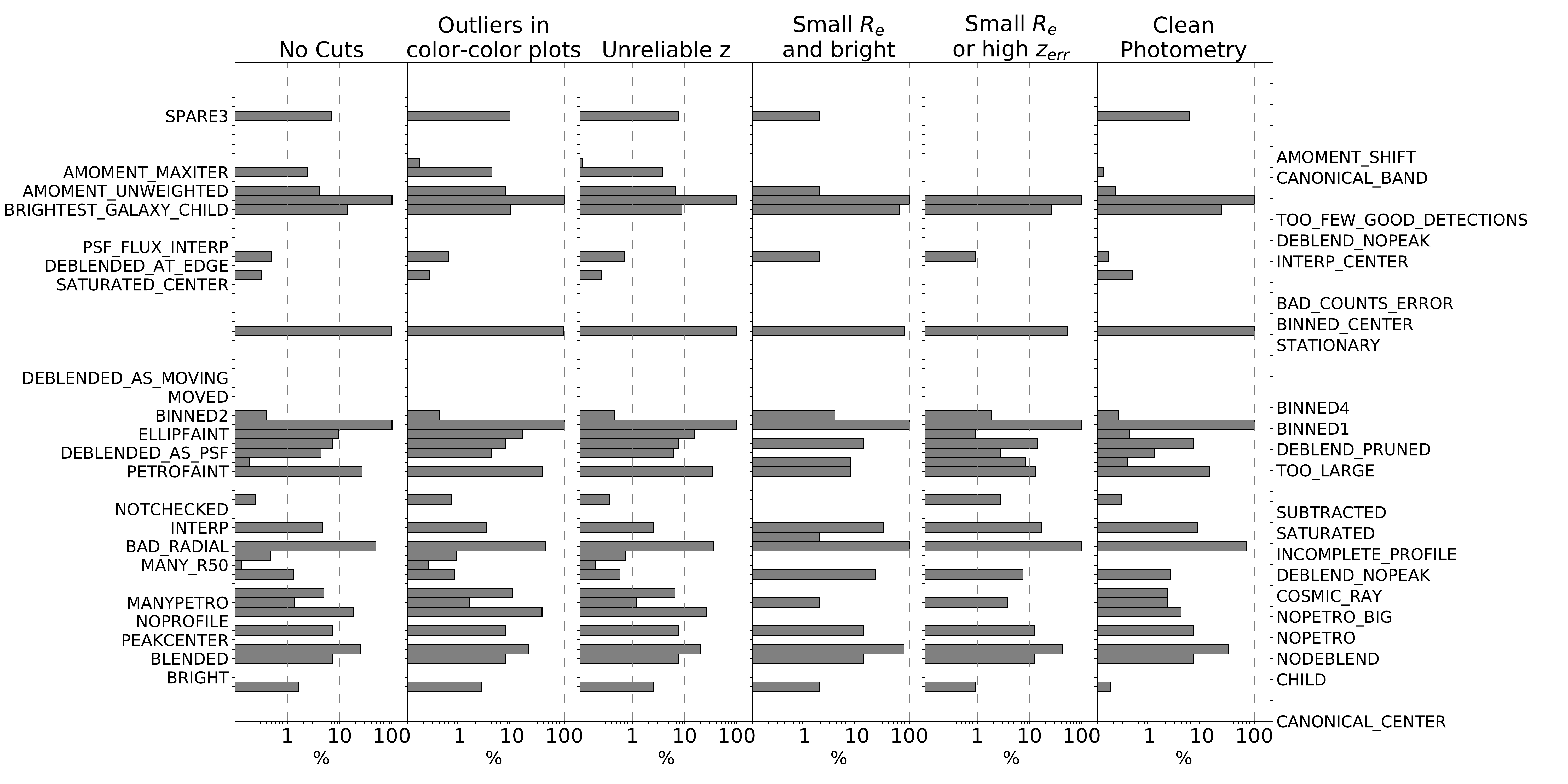}
\caption[Comparison with SDSS flags]{\textbf{Flags assigned to objects by the SDSS imaging pipeline.} Each bar counts the fraction of objects that have the corresponding flag set in the $r$-band. Each panel has different selection criteria indicated by each title. The final panel has all selection criteria applied.}
\label{fig:sdss_flags}
\end{figure*}

In \autoref{fig:sdss_flags}, we give a visual representation of the distribution of the flags in the $r$-band for different photometric selections. Each line corresponds to a particular flag and the length of each bar corresponds to the fraction of objects that have this flag set equal to \textit{True}. In all panels, \texttt{BINNED1} and \texttt{CANONICAL CENTER} are set to \textit{True} for all objects as a result of the \texttt{clean} shortcut. All flags which are deselected by the \texttt{clean} shortcut are absent from these objects. It is clear that while there are noticeable differences in each distribution, the overall distribution of flags does not change as a result of our photometric selections, confirming that our selections are indeed independent of the SDSS flags.\par

\begin{figure*}
\centering
\includegraphics[width=0.8\textwidth]{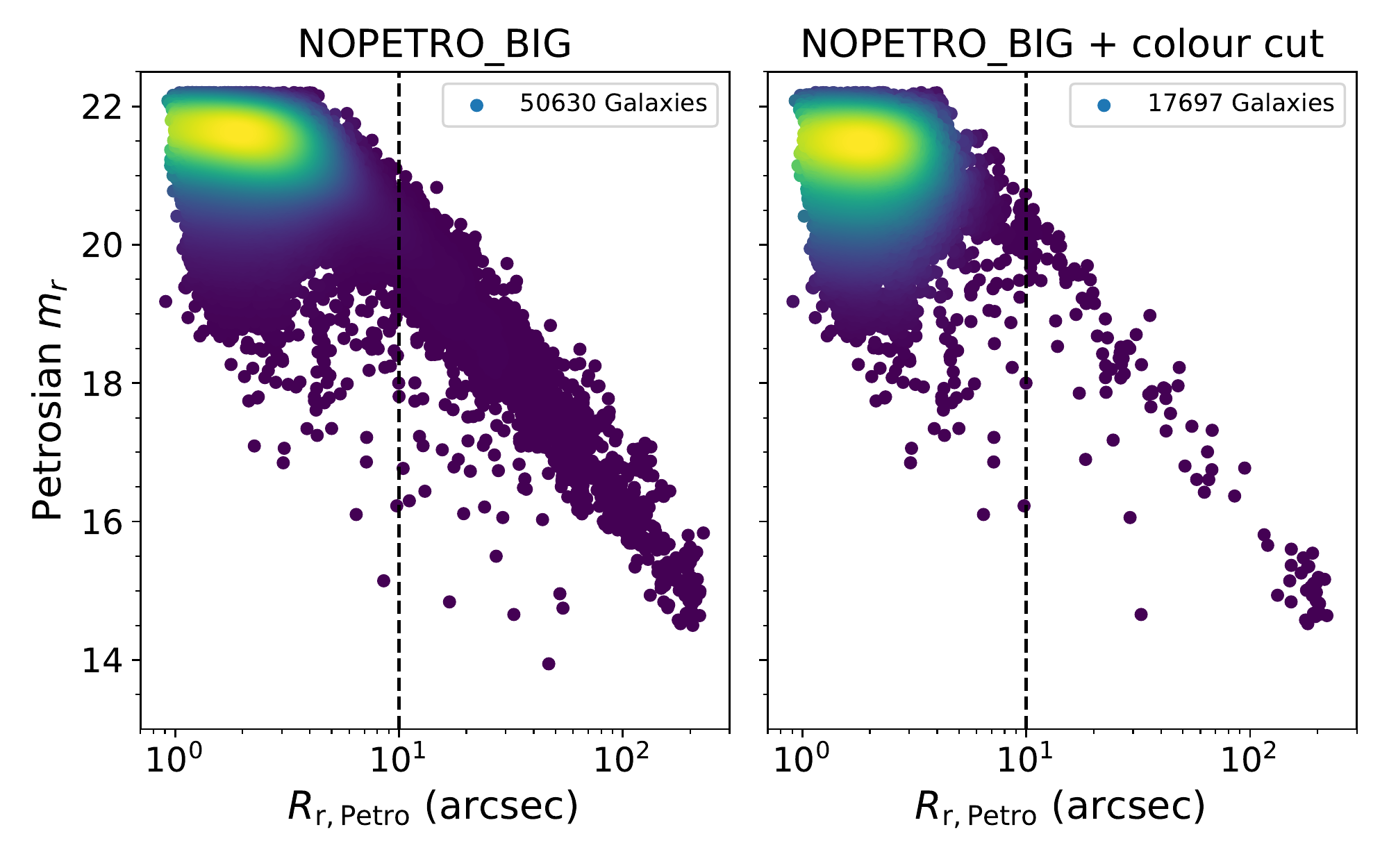}
\caption[$m_r$ versus $R_{\rm{r,Petro}}$]{\textbf{Left:} $m_r$ versus $R_{\rm{r,Petro}}$ for all galaxies in our 10x10 degree field with the \texttt{NOPETRO\_BIG} flag set in the $r$-band. The colour indicates the density of points. \textbf{Right:} The same except galaxies which fail to satisfy the colour-colour cut are not shown. About 35\% of the galaxies in the left plot remain in the right plot.}
\label{fig:nopetro_big}
\end{figure*}

As many of the flags are not critical for clean photometry, any differences between the panels for these flags can be ignored (such as \texttt{BAD RADIAL}). One flag we do check is \texttt{NOPETRO\_BIG}. This flag is set when the Petrosian radius is larger than the extracted radial profile, and can happen for faint objects with low signal-to-noise. In the top panel of \autoref{fig:nopetro_big}, we plot the Petrosian apparent $r$-band magnitude versus the Petrosian half-light radius for all objects in our 10 square degree test field that have \texttt{NOPETRO\_BIG} set equal to \textit{True}. The majority of objects are small and faint, but there is a tail which extends to large radii and bright magnitudes. This is because these objects are intrinsically faint and have low signal-to-noise, but the radial fit has become very extended, and hence the quoted magnitudes are overestimated as a result. About 98\% of all objects with $R_{\rm{r,Petro}}>10$ have the \texttt{NOPETRO\_BIG} flag set equal to \textit{True} confirming that the flag is very efficient at finding these objects. However, as stated, the vast majority of objects with this flag set to \textit{True} are small and faint, and hence are likely to be intrinsically small and faint. Hence, discarding all objects with this flag may not be desirable.\par
In the bottom panel of \autoref{fig:nopetro_big}, we plot the same quantities for the subset of objects in the top panel that satisfy our colour criteria. We find that using the colour criteria is almost as effective at removing the tail as \texttt{NOPETRO\_BIG}. However, our criteria leave a clean sample of galaxies at the faint end, although only about a third of the objects remain. Note that the factor of three decrease between the top and bottom panels does \textit{not} reflect the overall reduction due to the colour cut, as the majority of these objects are fainter than $20-21$ mag, which are more affected than brighter magnitudes, and these objects all have \texttt{NOPETRO\_BIG} set, which is not true for all objects.\par
As a result, we do not apply the \texttt{NOPETRO\_BIG} flag to exclude objects in the tail as our colour criteria take care of this with reasonable accuracy. Just to be complete, we exclude all objects larger than $R_{\rm{r,Petro}}=10$ arcsec which will not in any way discard the general population (see \autoref{fig:re_cuts}).

\subsection{Summary}
Here we summarise the criteria we use to obtain a clean photometric sample. They are given alongside the reasons for applying those criteria.

\begin{enumerate}
\item \texttt{type = 3} corresponding to `galaxy' as determined by the SDSS pipeline. This ensures that point-like stars are not included.
\item Clean photometry tagged with \texttt{clean = 1} and \texttt{calibStatus\_r \& 1) $\neq$ 0}. These are recommended to select clean photometry.
\item Five colour criteria given in \cref{eq:colour_criteria}. This is to exclude some stars and other unusual objects (such as patches of spiral arms/cosmic rays etc.).
\item $R_{\rm{r,Petro}} \leq 10$. This is to discard any objects where the fit in the $r$-band has failed and have not been removed by the colour criteria (see \autoref{fig:nopetro_big}).
\item For $m_r<16$, we require that $R_{\rm{petro,r}} \geq 1.5$. This is to exclude nearby K stars which have not been classified as stars by the imaging pipeline due to their extended appearance.
\item For $16 \leq m_r < 19$, we require that $R_{\rm{petro,r}} \geq 1.5$ \textit{or} $\Delta z_{\rm{phot}} \leq 0.075$. This is also to exclude a small fraction of stars as in the previous cut.
\end{enumerate}

Our only other criteria we use are \texttt{photoErrorClass} = [-1, 1, 2, 3] \& $\texttt{nnCount} \geq 95$. We exclude galaxies if they do not satisfy this criteria as their redshift estimates are too uncertain. We find that these galaxies constitute about 10\% of the population at $m_r\approx21$ mag and 50\% at $m_r \approx 21.5$ mag. 
\section{Compiling the MaNGA group catalogue}
In the previous section, we cleaned the SDSS photometric catalogue using empirical criteria. By combining this catalogue with the NSA, we have a magnitude limited sample of galaxies at all redshifts within MaNGA. 
\label{sec:samples}
\subsection{MaNGA data}
In this section, we use data from MPL-7 (compared with MPL-5 in \citealp{graham2018angular}) which contains data for 4597 unique galaxies and was released internally to the collaboration in Spring 2018. MPL-7 is identical to the MaNGA content of SDSS DR15 \citep{aguado2018fiftheenth}. As in \cite{graham2018angular}, we make use of stellar kinematics provided by the Data Analysis Pipeline (DAP; \citealp{westfall2019pipeline}) which is a fully-automated procedure that uses pPXF \citep{cappellari2004parametric, cappellari2017improving} to derive the stellar and gas kinematics from the stellar spectra. We obtain measurements for stellar mass, $\epsilon$, $\lambda_{R_e}$ etc. for all MaNGA galaxies using the method described in detail in sec. 3 of \citealp{graham2018angular}. (For a complete list of quantities that we extract, see Table 2 of \citealp{graham2018angular}.)

\subsection{A new mass-luminosity relation}
\label{sec:photometry_mass}
Ultimately our aim is to produce a complete catalogue which includes classifications of all candidate slow rotators. For this classification we will not use kinematics alone, but also stellar mass, because it was shown that genuine dry merger relics are essentially absent below a characteristic stellar mass $M_{\rm{crit}} = 2 \times 10^{11} \textrm{ M}_{\odot}$ (see \citetalias{cappellari2016structure} for a review of this evidence). For this reason we need to estimate well calibrated stellar masses for each galaxy.\par
For the NSA, the stellar mass is calculated by multiplying the rest-frame absolute luminosities, calculated using the K-correction, with a colour-dependent $M/L$ ratio \citep{blanton2007k}. However, we have no such estimate for the photometric catalogue. Dynamical masses correlate linearly with absolute $K_s$-band magnitudes \citep{cappellari2013effect} obtained from the 2MASS Extended Source Catalogue (XSC; \citealp{skrutskie2006two}), but we do not have access to the XSC photometry for all sources as the sensitivity of 2MASS is not the same as the SDSS. We do however have $r$-band apparent magnitudes for all sources from the SDSS combined catalogue, and so we fit a new relation using absolute luminosities in the $r$-band. Of course, in general, galaxies of a given stellar mass can have a range of $r$-band luminosities, as their mass-to-light ratio (M/L) varies with $\sigma$ rather than mass alone. However this M/L variation is dramatically reduced around and above $M_{\rm{crit}}$ (e.g. fig. 22 of \citetalias{cappellari2016structure}), which is the mass we use for selecting massive slow rotators. This ensures that although our stellar masses will not be very acccurate at the lower masses, they will be quite accurate for massive galaxies, ensuring a reliable rejection of non-massive galaxies.\par
\begin{figure*}
\centering
\includegraphics[width=\textwidth]{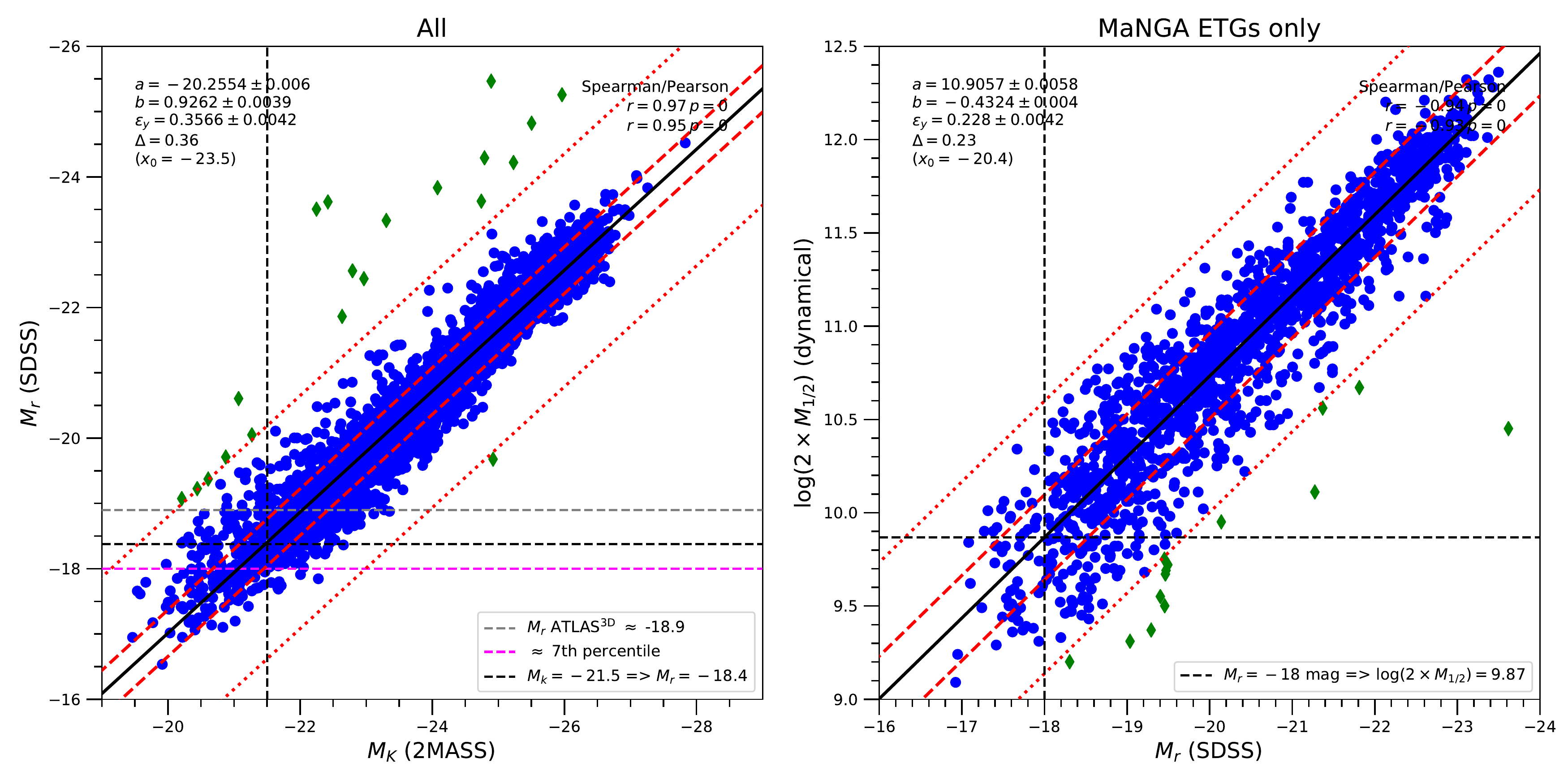}
\caption[Dynamical mass-luminosity relation]{\textbf{Dynamical mass-luminosity relation.} \textit{Left}: We plot the luminosity in the 2MASS $K_s$ band and the SDSS $r$-band for 3813 MaNGA galaxies of all morphologies which have 2MASS XSC photometry. \textit{Right}: We plot the dynamical mass-luminosity relationship for 1576 ETGs from MPL-5. $M_{1/2}$ is the enclosed mass within a three-dimensional half-light radius from dynamical models taken from \protect\cite{li2018population}.}
\label{fig:r_band_stellar_mass}
\end{figure*}

Before we fit $M_r$ to the stellar mass, we first find the $M_r$ corresponding to the ATLAS$^{\rm{3D}}$ limit $M_{K_s}=-21.5$ so that we are comparable to the ATLAS$^{\rm{3D}}$ survey, which represents the benchmark of IFU surveys. To do this, we use \texttt{lts\_linefit}\footnote{\label{Capwebpage}\url{http://purl.org/cappellari/software}} described in section 3.2 of \cite{cappellari2013atlas3db} to fit $M_r$ as a function of $M_{K_s}$ for 3813 galaxies from MPL-7 which are in the 2MASS XSC (see left panel of \autoref{fig:r_band_stellar_mass}). We find that there is an almost one-to-one correlation between the two luminosities with only a handful of outliers outside of $5\Delta$. In the ATLAS$^{\rm{3D}}$ survey paper, $M_{K_s}=-21.5$ was found to correspond approximately to $M_r = -18.9$ \citep{cappellari2011atlas3d}. In our relation, we find that $M_{K_s}=-21.5$ corresponds to $M_r \approx -18.4$ (we have about a factor 10 more galaxies). However, we find that about 13\% of MaNGA galaxies lie below this limit, and so we choose a fainter limit of $M_r=-18$, below which only $\sim8\%$ of MaNGA galaxies lie.\par
We fit a new mass-luminosity relation using the half-light mass estimated from realistic galaxy models made using the Jeans Anisotropic Modelling (JAM) method of \cite{cappellari2008measuring} for 1576 ETGs in MaNGA tabulated in Table A1 of \cite{li2018population}. We correlate the $M_r$ luminosities to the $M_{\rm{JAM}}$ stellar masses, as was done for the $K_s$-band by \cite{cappellari2013effect}. The advantage of using JAM dynamical mass estimates instead of stellar population values to transform our $r$-band luminosities is that the former do not suffer from the uncertain normalisation of about a factor two which is likely due to the systematic variation of the initial mass function with mass \citep{van2010stars,cappellari2012variation}.\par
This relation is shown in the right panel of \autoref{fig:r_band_stellar_mass}. We compare our slope of -0.4324 with that of \cite{cappellari2013effect} (-0.4449) and find that they are very similar. There is an offset between the different bands of about 3 mag, but this is purely due to the offset seen in the top panel of \autoref{fig:r_band_stellar_mass}. From this relation, we find that we are complete down to a stellar mass $M\approx 7.4\times 10^{9} \textrm{ M}_{\odot}$. If instead we use spirals, we find a slightly shallower slope (-0.355), nevertheless we apply the ETG relation to all galaxies.\par
\begin{figure}
\centering
\includegraphics[width=0.49\textwidth]{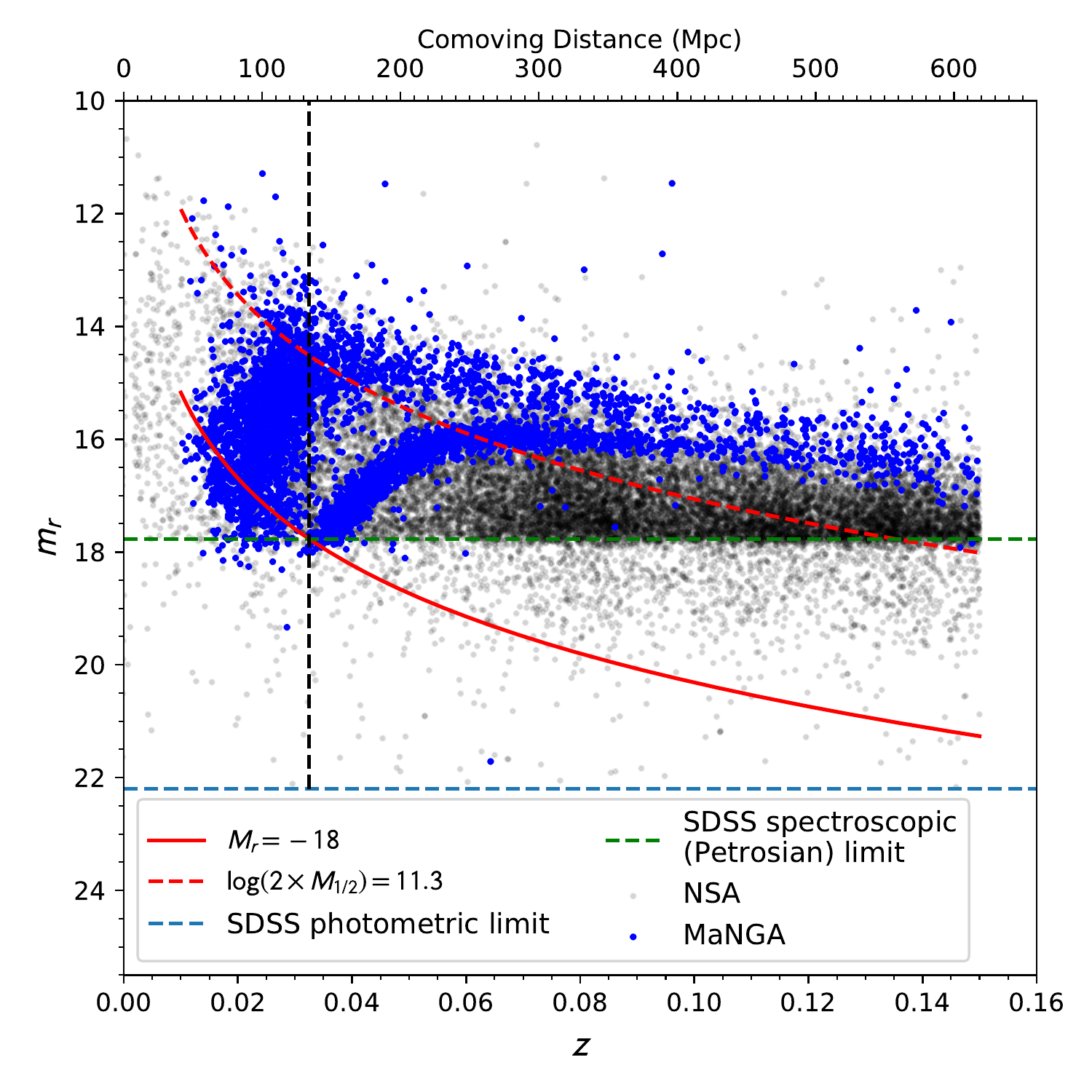}
\caption[NSA completeness as a function of redshift]{\textbf{NSA completeness as a function of redshift.} We plot the apparent magnitude and $z$ distributions of NSA (black) and MaNGA (blue) galaxies. The primary and secondary MaNGA samples are visible as two separate streams. The completeness limits for the NSA and the photometric sample are shown as green and blue dashed lines respectively. Finally, we plot the apparent magnitude corresponding to an absolute magnitude of -18 mag (solid red), as well as the absolute magnitude corresponding to the critical mass for SRs from the bottom panel of \autoref{fig:r_band_stellar_mass} (dashed red). The vertical dashed line indicates the approximate redshift ($z \approx 0.0325$) below which the NSA is complete down to $M_r=-18$.} 
\label{fig:fluxes}
\end{figure}

In Figure \ref{fig:fluxes}, we show the distribution of galaxy fluxes with redshift for $0.01 \lesssim z \leq 0.15$. Two sequences can be seen in the MaNGA distribution (blue) corresponding to the Primary and Secondary samples. The black points are galaxies from the NSA which are brighter than $m_r=17.77$ mag. Beyond $z\approx0.0325$, the NSA is not complete in the luminosity function down to $M_r=-18$ mag. The combined photometric and NSA catalogue is complete out to $z=0.15$.

\subsection{Finding neighbours of MaNGA galaxies}
\label{sec:group_finder}
As we are interested in the environment of MaNGA galaxies specifically, we use the MaNGA galaxies as our starting point. For this we need to find a volume-limited mass-selected sub-sample of SDSS galaxies which includes the MaNGA galaxies. We repeat the method outlined below once for each MaNGA galaxy in our sample, meaning we find one set of neighbours for each MaNGA galaxy. We refer to each starting MaNGA galaxy as the ``host'' galaxy of its set of neighbours to differentiate it from other potential MaNGA galaxies that may be enclosed within the same set.

\subsubsection{Excluding interlopers using redshift}
\label{sec:interlopers_redshift}
For each MaNGA galaxy, we select galaxies from the catalogue which lie within a cylinder with radius equal to 10 comoving Mpc centred on the MaNGA galaxy. We remove photometric objects that coincide with stars within 2 arcsec. For duplicate galaxies i.e. galaxies that appear in both the NSA and the photometric catalogue, we discard the entry from the photometric catalogue as the spectroscopic redshift is much more accurate. The remaining objects constitute the neighbouring galaxies. The height $h$ of the cylinder along the line of sight is by default $600 \textrm{ km s}^{-1}$ ($\Delta V = 300 \textrm{ km s}^{-1}$) to be consistent with \cite{cappellari2011atlas3db}. However, we extend this up to $\Delta V = 3000 \textrm{ km s}^{-1}$ for MaNGA galaxies that lie in large clusters with high velocity dispersion.\par
The peculiar velocity $\Delta V_{\rm{neigh}}$ between two galaxies assumed to be at the same cosmological redshift $z_{\rm{cosm}}$ is (see Eq. (9) of \citealp{hogg1999distance} or Eq. 10 of \citealp{cappellari2017improving})
\begin{equation}
\Delta V_{\rm{neigh}} = V_{\rm{neigh}} - V_{\rm{MaNGA}} = \Bigg( \frac{z_{\rm{neigh}} - z_{\rm{MaNGA}}}{1 + z_{\rm{MaNGA}}} \Bigg) c,
\label{eq:old_v_calc}
\end{equation}
where $c$ is the speed of light in km s$^{-1}$. By comparing the difference between $z_{\rm{neigh}}$ and $z_{\rm{MaNGA}}$ with $1 + z_{\rm{MaNGA}}$, \cref{eq:old_v_calc} states that the velocity difference between two redshifts depends on the cosmological redshift. \citealp{baldry2018redshifts} makes the case that $\ln(z)$ is a more natural distance indicator because a velocity measurement is effectively a shift on a logarithimic wavelength scale, as galaxy spectra are binned linearly with $\ln(\lambda)$ (e.g. \citealp{tonry1979redshifts, cappellari2017improving}). If we instead define $\Delta V_{\rm{neigh}}$ as
\begin{equation}
\Delta V_{\rm{neigh}} = V_{\rm{neigh}} - V_{\rm{MaNGA}} = [ \ln(1+z_{\rm{neigh}}) - \ln(1+z_{\rm{MaNGA}}) ] c\textrm{,}
\label{eq:new_v_calc}
\end{equation}
then the cosmological and peculiar velocities add linearly, without the need to ``deredshift" from the cosmological redshift. When the peculiar velocities are of order 100 km s$^{-1}$, then \cref{eq:old_v_calc} and 
\cref{eq:new_v_calc} are equivalent to within 1 km s$^{-1}$ (see fig. 1 of \citealp{baldry2018redshifts}). As we are free to use either definition, we choose to use \cref{eq:new_v_calc} as it is mathematically more elegant than \cref{eq:old_v_calc}.\par
For the spectroscopic neighbours, we could simply check that $|\Delta V_{\rm{neigh}}| \leq \Delta V_{\rm{cyl}}$. However, this would not work for photometric galaxies as we need to check if the photometric redshift errors, converted to velocity, overlap with the cylinder. Hence, we use a more general function,
\begin{numcases}{f_{\rm{overlap}}(V_{\rm{min}},V_{\rm{max}},h)=}
  1, & if $V_{\rm{min}} \leq h/2$ and $V_{\rm{max}} \geq -h/2$ \nonumber \\
  0, & otherwise
\label{eq:overlap}
\end{numcases}
where $V_{\rm{min}}$ and $V_{\rm{max}}$ are the lower and upper velocity limits calculated using \cref{eq:overlap} for each galaxy and $f_{\rm{overlap}}=1$ corresponds to an overlapping galaxy. To satisfy $f_{\rm{overlap}}=1$, a galaxy needs only to overlap with the cylinder and not $V=0 \textrm{ km s}^{-1}$ corresponding to the MaNGA galaxy itself. For galaxies in the NSA, $V_{\rm{min}}$ is equal to $V_{\rm{max}}$ within our precision of $\sim30 \textrm{ km s}^{-1}$. In contrast, photometric galaxies can easily have velocity errors that are much larger than the height of the cylinder. However, we have no reason to exclude these galaxies on the basis of large errors in redshift. If we were to estimate the local density using only the spectroscopic catalogue, we would grossly underestimate the number density of the high redshift galaxies in our sample, which are the most massive and most likely to be SRs.\par 
Here we quantify the probability that a given photometric galaxy will actually lie within a given cylinder. We assume that the errors on the photometric redshifts provided in \cite{beck2017redshift} are reliable. We also assume that the likelihood of a galaxy lying within its error bounds is uniform. In fact, this likelihood is a Gaussian centred on the most likely value, but as our discussion is only approximate, this assumption is not critical. For a cylinder with $\Delta V = 300 \textrm{ km s}^{-1}$, the probability of overlap is at best about 20\%\footnote{For a minimum $z_{\rm{err}}=0.005$ for photometric galaxies, $P(\rm{overlap})\approx 0.001/0.005=20\%$ assuming maximal overlap and $\Delta V/c\approx 0.001$.} but could be as low as $\sim0.01\%$\footnote{For $z_{\rm{err}}=0.2$ for photometric galaxies, $P(\rm{overlap})\approx 0.001*0.01/0.2=0.005\%$ assuming 1\% of the cylinder overlaps and $\Delta V/c\approx 0.001$.}. It is possible that for the few groups where the cylinder is extended in velocity, the probability of overlap can equal 100\%. However, the probability of \textit{multiple} photometric galaxies lying in a single cylinder rapidly approaches zero as the number of photometric galaxies increases.\par

\subsubsection{Excluding interlopers using luminosity}
\label{sec:interlope_lum}
\begin{figure*}
\centering
\includegraphics[width=0.95\textwidth]{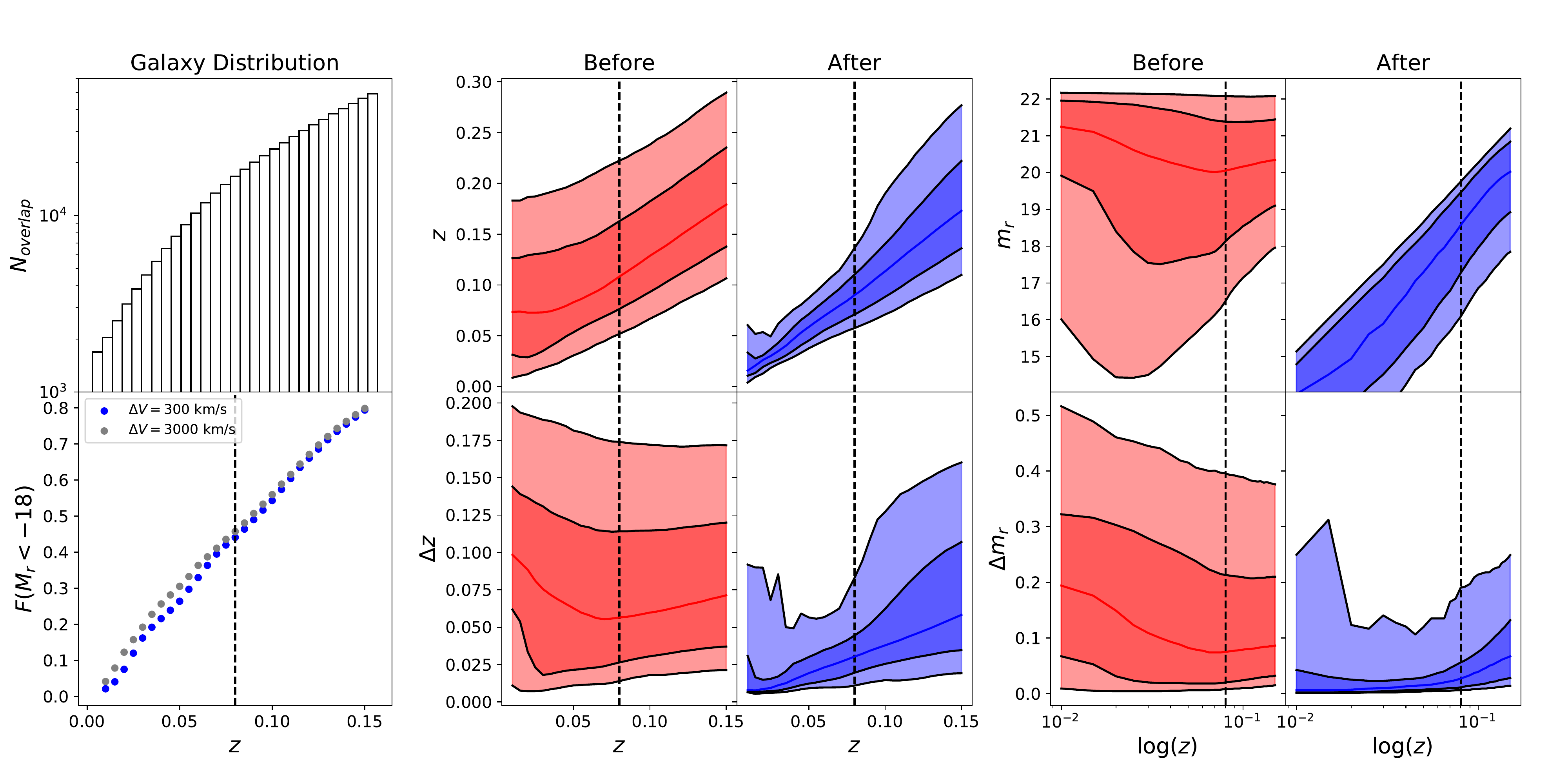}
\caption[Applying the luminosity cut]{\textbf{Applying the luminosity cut.} \textbf{Left}: We plot $F(M_r \leq -18)$ as a function of redshift for galaxies in the photometric sample. At each redshift $z$, the fraction is calculated as the number of galaxies that overlap with a redshift cylinder with height $z=0.002$ centred on $z$ \textit{and} satisfy our minimum absolute magnitude $M_r=-18$ assuming they are at redshift $z$ divided by the total number of galaxies that overlap at redshift $z$. Hence, the number of galaxies in each $z$ bin is not constant. \textbf{Right}: In the upper left set of two panels, we plot the redshift distribution of galaxies that overlap the redshift cylinder centred at each $z$ before (red, left) and after (blue, right) the luminosity cut is applied. The median is shown as the thick coloured line and the inner $1\sigma$ and $2\sigma$ percentiles are shown as shaded regions. The upper right set of panels is the same for the apparent magnitude distribution. The bottom sets of panels illustrate the corresponding error distributions. For the right set of plots, we show the $x$-axis as a logarithmic scale as the apparent magnitude increases as a function of $\log(z)$. In all panels, the upper redshift limit of the complete sample is indicated with a dashed vertical line (see \autoref{sec:maximum_z}).}
\label{fig:frac_mag_cut}
\end{figure*}

Clearly, it makes little sense to include galaxies with only a photometric redshift with large errors that happens to overlap with the cylinder, especially if they are at high redshift. We can improve the selection and further eliminate interlopers by assuming that if they are at the redshift of the MaNGA galaxy, their apparent magnitude should be such that their absolute magnitude satisfies our minimum luminosity of $M_r=-18$ mag. In essence, we move each galaxy from its recorded photometric redshift to the redshift of the MaNGA galaxy, which it must be at if it is to be bound to the same dark matter halo. Galaxies at high redshift that are brought forward will decrease in luminosity (and stellar mass), and vice versa. To quantify the effectiveness of this extra quality control step, we select $N_{\rm{overlap}}$ photometric galaxies that overlap with cylinders spaced at regular redshift intervals between $z=0.01$ and $z=0.15$. At each redshift, we calculate the fraction of $N_{\rm{overlap}}$ galaxies that survive the luminosity cut i.e. $M_r \leq -18$ mag assuming they are at the redshift of the cylinder. We find that this fraction increases linearly with redshift regardless of the height of cylinder (see lower left panel of \autoref{fig:frac_mag_cut}). As expected, increasing $h$ from $\sim 300\textrm{ km s}^{-1}$ to $\sim 3000\textrm{ km s}^{-1}$ only increases the fraction of surviving galaxies by at most $\sim5\%$. Only about 5\% of galaxies survive the luminosity cut at $z\sim0.01$ compared with $\sim80\%$ at $z\sim0.15$.\par
We also compare the redshift and apparent magnitude distributions (and their corresponding errors) as a function of redshift before and after applying the luminosity cut (see middle and right panels right of \autoref{fig:frac_mag_cut}). We find that by applying the luminosity cut, we automatically select galaxies which (1) are lower in redshift and (2) are brighter and so therefore they have (3) smaller errors in redshift and (4) in apparent magnitude. After the luminosity cut, the median $z$ of surviving galaxies increases linearly with $z$, and the error also generally increases with $z$. The apparent magnitude increases linearly with $\log(z)$ and the magnitude errors remain roughly constant with a sight upturn beyond about $z\sim0.1$. Hence, applying the luminosity cut is crucial for us to obtain a cleaner sample of galaxies than would be achieved with just the velocity cut.

\subsection{Constructing the group catalogue}
\label{sec:groups}
For each MaNGA galaxy, we find the surviving neighbours that satisfy all our criteria i.e. the clean photometric selection as well as the velocity and luminosity cuts. We then run the group finder algorithm \texttt{TD-ENCLOSER}, presented in \citetalias{graham2019atechnical}, using each set of surviving neighbours as input. We then find the subset of galaxies which \textit{encloses} each MaNGA galaxy i.e. we find all galaxies that belong to the same density peak as the MaNGA galaxy. These galaxies constitute a \textit{set of neighbours} which are ``hosted'' by the MaNGA galaxy. We do not call these \textit{groups} (yet) for the following reason. As some MaNGA galaxies will be physically close to other MaNGA galaxies (especially at low redshift), some of these \textit{sets} will be duplicates of the same intrinsic \textit{groups}. We do not know in advance which sets will be duplicates. We account for this duplicity in \citetalias{graham2019atechnical} where we present a simple procedure to remove duplicates sets. However, we only apply this procedure after completing the task outlined below, namely the visual classification of galaxies, as we do not know which duplicates will be removed.\par
\subsubsection{Environmental measures}
\label{sec:params}
At this point, we have a group catalogue with which we can study galaxy environment. In this subsection, we introduce the parameters we will use for our science analysis presented in \citetalias{graham2019cenvironment}.\par
There are a variety of different methods found in the literature that are designed to quantify galaxy environment. \cite{muldrew2012environment} compared two broad groups of measures, one based on the distance to the $n$th nearest neighbour, and another based on the number of galaxies enclosed within a fixed aperture. Their analysis concluded that the methods based on the $n$th nearest neighbour are better probes of the local density, while aperture based methods are more sensitive to the large scale density. Moreover, \cite{cappellari2011atlas3db} found that galaxy morphology is affected more by the projected number density $\Sigma_3$ compared to a more distant measure such as $\Sigma_{10}$. Nearest neighbour methods have also been used by a number of other studies to study galaxy morphology as a function of environment (e.g. \citealp{baldry2006environment, park2009environment, tonnesen2012environment, etherington2015environment, goddard2017environment, brough2017kinematic, greene2017kinematic, schaefer2017environment, lee2018environment}). Here we focus on the projected number density $\Sigma_3 = 3/(\pi D_3^2)$ where $D_3$ is the projected comoving distance in Mpc of the third nearest neighbour to a galaxy.\par
Another commonly used parameter is the mass of the dark matter halo that hosts a group of galaxies. The halo mass is the only halo property to correlate with local galaxy density to the first order \citep{lemson1999environment, haas2012environment} and can be measured accurately in dark matter simulations. We do not have access to halo masses, but we take advantage of the halo mass-richness relation (e.g. \citealp{becker2007clusters, rozo2009clusters, shen2014galaxies}) which allows the group richness $\mathcal{N}$ to be used as a proxy. Finally, for each galaxy, we measure the projected comoving distance to the galaxy at the centre of the group as determined by \texttt{TD-ENCLOSER}, $D_{\rm{cen}}$. We do not choose the most massive galaxy to be the central galaxy. Instead, we designate the galaxy that lies at the peak of the density field to be the central galaxy, so that we do not bias the measurement in any way. The ETG fraction is known to correlate with the cluster-centric radius \citep{whitmore1991density, whitmore1993morphological, hansen2009clusters}, and hence it would be logical to expect that the fraction of massive SRs would also vary as a function of $D_{\rm{cen}}$. As we also do not have virial radii to hand, we consider both the physical group-centric radius $D_{\rm{cen}}$ and the group-centric radius normalised to the 90th percentile, which is less affected by outliers than the 100th percentile.\par
Although it has been established that the abundance of SRs is larger at the centre of clusters, \cite{cappellari2011atlas3db} showed that the SR fraction F(SR) is not a simple function of $\Sigma_3$, unlike for the classic T-$\Sigma$ relation. In fact, outside Virgo, F(SR) was found to be nearly independent of $\Sigma_3$. The lack of a universal $f_{SR}-\Sigma_3$ relation was also found when studying different galaxy clusters. The clearest example can be seen in fig. 7 of \cite{scott2014distribution} which compares F(SR) for the Coma, Virgo and Fornax clusters as well as the ATLAS$^{\rm{3D}}$ field sample. At a fixed number density $\log(\Sigma_3) \approx 2$ Mpc$^{-2}$, SRs can either be absent or constitute one third of all ETGs. For the eight clusters in the SAMI cluster sample \citep{brough2017kinematic}, F(SR) can vary between clusters by up to 20\% for a given overdensity. By averaging F(SR) profiles from multiple clusters to produce a single kT-$\Sigma$ relation, we risk washing out the clear trends seen in the F(SR) profiles for individual clusters. In light of this, we define the \textit{relative} number density $\Sigma_3^{\rm{rel}}=\Sigma_3/\Sigma_3^{\rm{max}}$ where $\Sigma_3^{\rm{max}}$ is the maximum $\Sigma_3$ for a given cluster. (We maintain the subscript 3 for clarity, but this parameter can be calculated for any $n$ as $\Sigma_n^{\rm{rel}}$.) By this definition, a galaxy at the peak of the number density will have $\log(\Sigma_3^{\rm{rel}})=0$, and all other galaxies will have $\log(\Sigma_3^{\rm{rel}})<0$. In this way, all clusters are shifted to a common baseline. \cite{houghton2015density} already suggested that the fraction of elliptical galaxies depends on the relative local density within a cluster, but this has not yet been investigated for massive SRs. One possible interpretation of $\log(\Sigma_3^{\rm{rel}})$ is that it is a distance indicator, where the distance metric is number density. In this way, it is analogous to the projected distance $D_{\rm{cen}}$, which can also be considered as a relative estimator (i.e. relative to the central galaxy). However clusters are not spherically symmetric, but contain substructure, and for this reason $\Sigma_3^{\rm{rel}}$ and $D_{\rm{cen}}$ measure different slightly physical quantities.\par

\subsubsection{Achieving a volume-limited sample}
\label{sec:volume-limited}
While the MaNGA selection function is ideal for studying massive slow rotators as it allows for many massive galaxies to be observed, it is by design not a volume-limited sample. MaNGA does not have a dedicated cluster programme\footnote{There is an ancillary programme for the Coma cluster which we do not consider as it uses different observing strategies. See \citealp{gu2018coma} for a description and \citealp{albareti2016thirteenth} for a complete list of ancillary programmes.} and so this poses a challenge in how to interpret the kT-$\Sigma$ relation when the MaNGA sample only contains arbitrarily chosen members of multiple distinct clusters or groups. It would not make sense to obtain the kT-$\Sigma$ relation for a cluster with only partial observations with IFS, because our results will be affected by major selection biases that are difficult to correct for. Moreover, the results would not be comparable with single cluster studies or volume limited surveys. It is trivial to correct for the volume represented by every galaxy, producing a volume-weighted sample, by duplicating each MaNGA galaxy $N$ times where $N$ is inversely proportional to the size of the volume in which the galaxy would still be observed by MaNGA. However this will produce an artificial sample which does not reflect the real Universe. Small galaxy groups are much more numerous in the Universe than large groups or galaxy clusters. The mass-environment degeneracy means that MaNGA will observe galaxies in large groups more frequently than if the sample was a random, volume-limited sample. For example, we are very unlikely to observe low mass fast rotators in groups of 50 members because they are much more likely to be isolated or live in small groups.\par
A simple but non-trivial solution would be to take all galaxies in a cluster or group for which we do not have stellar kinematics and make an educated guess using the optical morphology as to the angular momentum classification (fast or slow rotator). This would work for spiral galaxies which form a parallel sequence to fast rotators and so the presence of spiral arms immediately rules out of the possibility of a galaxy being a SR. Unfortunately, there is a substantial lack of a one-to-one correspondence between the stellar kinematics and optical morphology of ETGs. In fact, two-thirds of galaxies classified as elliptical (i.e. spheroidal) are in fact fast rotators, which intrinsically are axisymmetric disk galaxies and would appear flattened if viewed edge on \citep{cappellari2011atlas3db}.\par
Thankfully, fast and slow rotators are not defined solely in terms of angular momentum, which is only obtainable with stellar kinematics, but also in terms of stellar mass and $\epsilon$ \citep{cappellari2013effect, cappellari2016structure}. The slow rotators are thought to be the relics of dry mergers and as such they are expected to be massive compared to fast rotators, which form by accretion of cold gas, and are also products of gas-rich (wet) mergers \citep{cappellari2016structure}. Observationally, galaxies classified as slow rotators using stellar kinematics have been shown to dominate above a critical mass of $2\times 10^{11} = 10^{11.3} \textrm{ M}_{\odot}$ as was initially pointed out by \cite{emsellem2011atlas3d} and reinforced by \cite{cappellari2013atlas3d} (see their fig. 14) in ATLAS$^{\rm{3D}}$. This characteristic was later confirmed by \cite{veale2016massive} with MASSIVE, \cite{greene2017kinematic} with MaNGA, \cite{brough2017kinematic} with SAMI as well as in \cite{graham2018angular}. Lower mass galaxies appear to be a qualitatively different class of objects and follow different formation channels (e.g. \citetalias{cappellari2016structure}). If a merger remnant is less massive than the critical mass, then its progenitors must not be fully quenched and in this case, the disk can remain stable and the merger remnant is a fast rotator ETG. For this reason it makes sense to exclude lower mass galaxies when studying massive slow rotators as a homogensous class. The dry merger channel is violent and disrupts the stellar disk as a result of the redistribution of angular momentum from within $1 R_e$ out into the galaxy halo. Hence, the slow rotator merger remnants are intrinsically round and so would appear round from any viewing angle. Therefore, any ETG that is less massive than $\log(M)=11.3$ \textit{or} appears to be flatter than $\epsilon = 0.4$ (an empirical demarcation line that was proposed by \citetalias{cappellari2016structure} and strongly confirmed in \citealp{graham2018angular}) cannot by definition be a genuine dry merger relic (i.e. a slow rotator).\par
Hence, we only have to visually classify the subset of galaxies that do satisfy  $M \geq M_{\rm{crit}}$ and $\epsilon<0.4$.

\subsubsection{Choosing a maximum redshift limit}
\label{sec:maximum_z}
Ideally, we would like to able to visually classify all the potential SRs in all sets of neighbours. However, there are three main reasons why we would not want to classify galaxies up to the maximum redshift of MaNGA ($z=0.15$):
\begin{enumerate}
\item At high redshifts ($z \gtrsim 0.1$), it is more difficult to classify galaxies visually as the resolution is poorer compared to lower redshifts.
\item As the completeness of the spectroscopic sample declines with redshift, the accuracy of the galaxy groups and hence the reliability of any statistical measurement also decreases (see \autoref{fig:fluxes}).
\item The MaNGA survey preferentially observes galaxies at lower redshifts ($z \lesssim 0.05$: see \autoref{fig:fluxes}), and hence the fraction of confirmed fast (or slow) rotators compared to the total will be greater for a redshift-limited subsample than for the whole sample.
\end{enumerate}
It is therefore necessary to specify an intermediate redshift between the MaNGA limits ($0.01 \leq z \leq 0.15$) below which we classify galaxies. However, an important constraint on this redshift is that the brightest and most massive galaxies in MaNGA are observed at the higher redshifts, and so the majority of \textit{confirmed} slow rotators will be at the higher redshifts (see \autoref{fig:fluxes}). Moreover, the reduction in sample size may mean that our number statistics become poorer as a result. In short, whichever redshift we choose to define as a upper limit for our subsample will be a compromise between resolution and accuracy of both the groups and the angular momentum (case for low $z$), and number statistics and number of confirmed slow rotators (case for high $z$). We find that roughly half ($\sim43\%$) of all \textit{confirmed} massive slow rotators in MaNGA lie below the redshift midpoint ($z=0.08$) and so we choose this redshift to define the upper limit of our volume-limited sample.\par
To summarise, we only classify the angular momentum of galaxies which satisfy $M \geq M_{\rm{crit}}$ and $\epsilon<0.4$ and lie in sets hosted by MaNGA galaxies which lie at a redshift $z\leq0.08$. We describe our classification method in detail in \autoref{sec:srcandidates}. We do not classify galaxies in sets that lie beyond $z=0.08$. In \autoref{sec:conclusions1}, we present the group catalogue. In our catalogue, we include \textit{MaNGA} galaxies in the redshift range $0.08 < z \leq 0.15$ for reference, but we do not use these galaxies for any science.

\subsubsection{Comparison with Yang et al. 2007 group catalogue}
Now that we have constructed our group catalogue, we compare our volume-limited sample catalogue with a well-established one. We choose the \cite{yang2007groups} catalogue updated to SDSS DR7 due to its popularity and the fact that it is based on SDSS (we use the one based on Petrosian magnitudes and sample II which supplements SDSS redshifts with additional redshifts from other surveys). It also extends to $z=0.09$ which is beyond the limit of our volume-limited sample. We first match galaxies to groups using \texttt{NYU\_ID}, and then calculate the richness for each group in the catalogue (as this is not provided). We then match each galaxy in our group catalogue to the one in the \cite{yang2007groups} catalogue using RA and Dec coordinates. We find that $\sim32\%$ of the galaxies in our catalogue do not have a match within 5 arcsec in the \cite{yang2007groups} catalogue, of which $\sim75\%$ only have a photometric redshift, and not a spectroscopic one. For each matched galaxy in our catalogue, we now have a richness for its enclosing group from this work ($\mathcal{N}^{\rm{TW}}$), and its enclosing group from \cite{yang2007groups} ($\mathcal{N}^{\rm{Yang}}$).\par
\begin{figure*}
\centering
\includegraphics[width=0.95\textwidth]{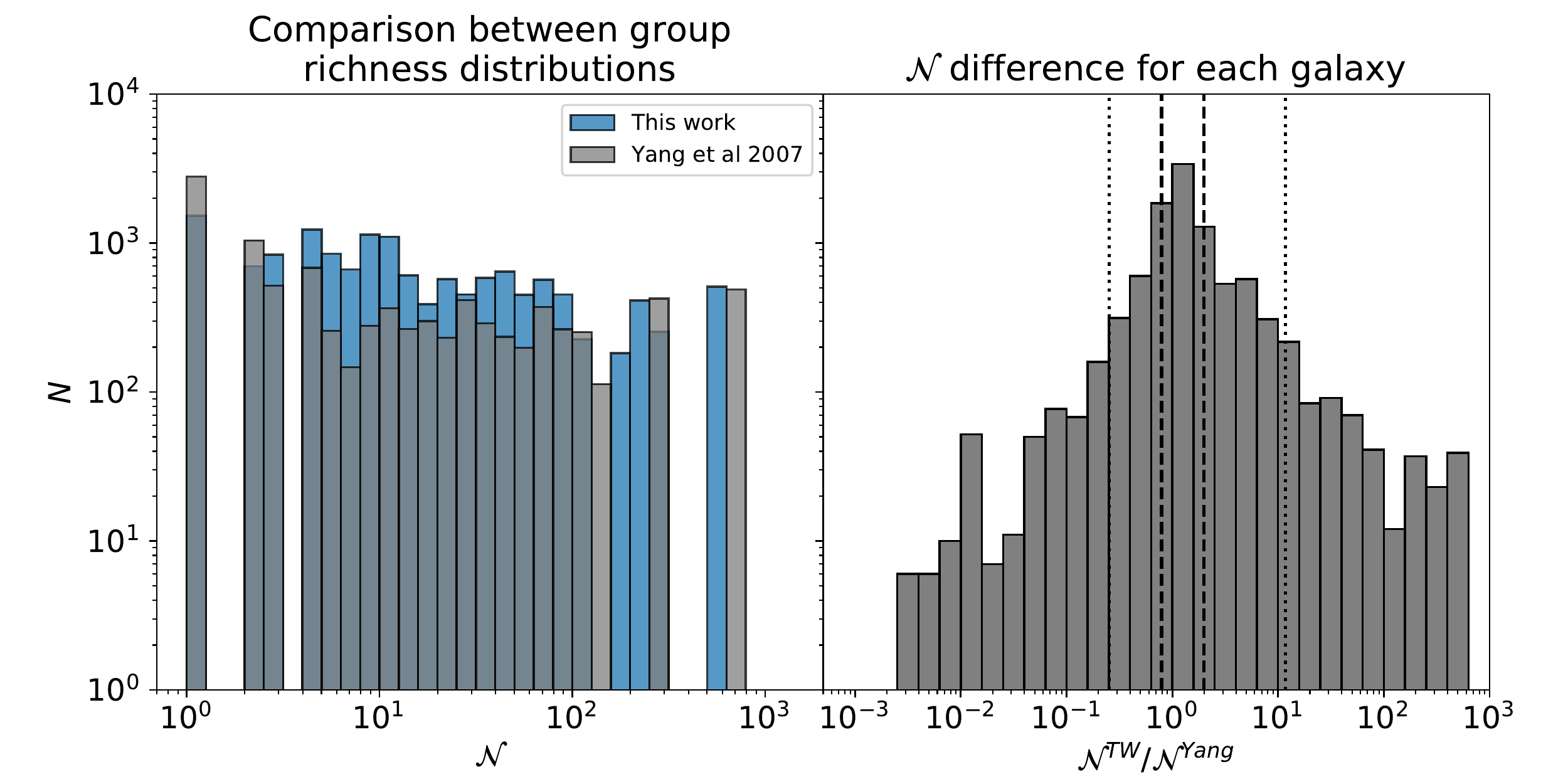}
\caption[Comparing group richness]{\textbf{Comparing group richness.} \textbf{Left}: For galaxies which are in our catalogue and have a group assigned in the \protect\cite{yang2007groups} DR7 catalogue, we plot the richness distribution from our catalogue in blue, and the richness distribution from the \protect\cite{yang2007groups} catalogue in grey. The bins are evenly spaced in log units. \textbf{Right}: For each galaxy present in both catalogues, we calculate $\mathcal{N}^{\rm{TW}}/\mathcal{N}^{\rm{Yang}}$ and compute the histogram for all galaxies. The dashed and dotted lines indicate the inner 50\% and 90\% respectively.}
\label{fig:richness_comp}
\end{figure*}
We plot the richness distributions for $\mathcal{N}^{\rm{TW}}$ and $\mathcal{N}^{\rm{Yang}}$ on the left hand side of \autoref{fig:richness_comp}. We find that while they are comparable overall, we tend to find larger groups more frequently than \cite{yang2007groups}, who conversely find more isolated galaxies than we do. Our group memberships are slightly biased towards larger groups as can be seen in the right hand side of \autoref{fig:richness_comp}, where we compare $\mathcal{N}^{\rm{TW}}-\mathcal{N}^{\rm{Yang}}$ for individual galaxies.\par
It is also useful to compare halo masses, as this is a standard parameter of group catalogues. To estimate the halo mass for the groups in our catalogue, we use the halo-richness relation of \cite{shen2014galaxies} (see their Eq. 2 and Figure 2) which estimates halo masses using abundance matching considering only galaxies with $M_r < -19.5$. We choose their relation to be consistent as it is based on the same group catalogue of \cite{yang2007groups}. Instead of using $M_r$, we estimate the stellar mass corresponding to $M_r = -19.5$ using the relation shown in \autoref{fig:r_band_stellar_mass} to be $10^{10.5} \textrm{ M}_{\odot}$. (\citealp{shen2014galaxies} have an alternate relation, given in their Eq. B3, which is calibrated for galaxies above this stellar mass but we do not expect any significant changes when using this relation due to its similarity to their Eq. 2.) The relation is a broken power-law where the halo mass $M_{\rm{halo}}^{\rm{Shen}}$ is calculated from the number of galaxies with $M_r<-19.5$, or in our case $M \geq 10^{10.5} \textrm{ M}_{\odot}$.\par
\begin{figure*}
\centering
\includegraphics[width=0.49\textwidth]{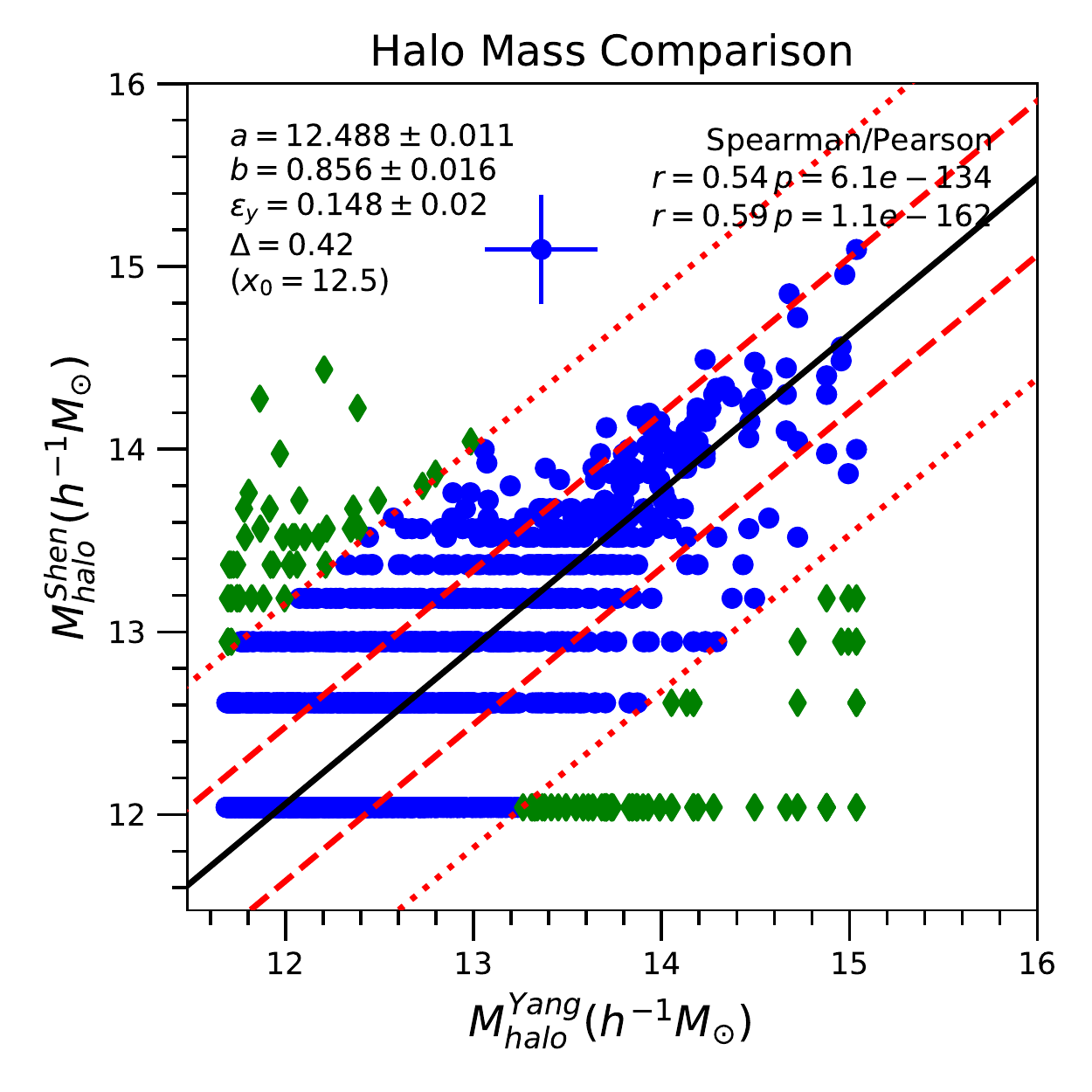}
\caption[Comparing cluster mass]{\textbf{Comparing cluster mass.} On the x-axis, we plot the halo mass from the \protect\cite{yang2007groups} catalogue, $M_{\rm{halo}}^{\rm{Yang}}$, and on the y-axis, we plot the halo mass for our group catalogue estimated using Eq. 2 of \protect\cite{shen2014galaxies}. As with similar plots in this thesis, the best fitting relation is calculated using \texttt{LTS\_LINEFIT}. The errorbars along each axis are shown in the upper middle of the plot (where no real galaxy lies).}
\label{fig:halo_mass_comp}
\end{figure*}

We compare our estimated halo mass with the halo mass given in the \cite{yang2007groups} catalogue ($M_{\rm{halo}}^{\rm{Yang}}$) in \autoref{fig:halo_mass_comp}. We allow for the error in log units to be 0.3 for both catalogues otherwise the fit is not successful, likely due to the discrete nature of our halo masses (as they are calculated from a discrete group richness). We measure the slope and scatter of the distribution using \texttt{LTS\_LINEFIT}. We find that while there is a large amount of scatter, the slope is not far from one, and the intercept at the pivot is almost identical to the pivot value itself, suggesting that there is little to no offset between the two halo mass estimates. The broad agreement between our estimated halo masses and those of \cite{yang2007groups} gives confidence that the accuracy of our group finder algorithm and method is sufficient.\par
Finally, we compare which groups and clusters we have in common with \cite{yang2007groups}. We find that while very few groups match up entirely, about 2000 ($\sim 55\%$) of our groups (including isolated galaxies) have all their members present in the \cite{yang2007groups} catalogue split between multiple groups, although almost all of these have $\mathcal{N}<10$ in our catalogue. Of the largest 10 groups in our catalogue which we find to have at least about 90 galaxies each, seven have at least 50\% of the members present in the \cite{yang2007groups} catalogue, while the remaining three have between 30\% and 50\% present (again split between different groups in the \cite{yang2007groups} catalogue). About 100 groups with $\mathcal{N} \lesssim 15$ do not have any members present in the \cite{yang2007groups} catalogue. All groups with $\mathcal{N}$ in our catalogue have at least 20\% of their members in the \cite{yang2007groups} catalogue. Comparing the content of two group catalogues is challenging, as even if two overlapping groups have the same $\mathcal{N}$ in both catalogues, the member galaxies may not be the same set of galaxies. 
\section{Selecting the slow rotator candidates}
\label{sec:srcandidates}
Before we classify galaxies without stellar kinematics as fast or slow rotators, it is crucial to understand how accurately we can make the classification. \cite{cappellari2011atlas3db} found that only one third of elliptical galaxies are genuinely spheroidal (slow rotators) and that two thirds are in fact misclassified face-on disks (fast rotators). However, they did not take into account the stellar mass or ellipticity criteria that we enforce here. Can we use the extra criteria to improve the accuracy rate for slow rotators? The MaNGA sample is ideally suited for answering this question as it contains a large number of high-mass galaxies compared to a volume limited sample of the same sample size \citep{wake2017sdss}.
\subsection{Visual classification using morphological features}
\label{sec:visual_class}
We answer the above question by guessing the angular momentum classification of 581 MaNGA galaxies that satisfy the SR criteria ($\log(M)\geq11.3$ and $\epsilon < 0.4$), and compare our guesses to the true classifications obtained using $\lambda_{R_e}$. In our test, we look out for known characteristic signatures: SRs are massive ellipticals with a bright bulge and an extended stellar halo (e.g. \citealp{schombert1985brightest}). All disk-related features such as bars and rings are absent, although they can possess shells which are indicative of a recent dry merger event. In contrast, the majority of FRs exhibit disk-like features with only a handful appearing morphologically indistinct from SRs.\par
\begin{figure*}
\centering
\includegraphics[page=1,width=\textwidth]{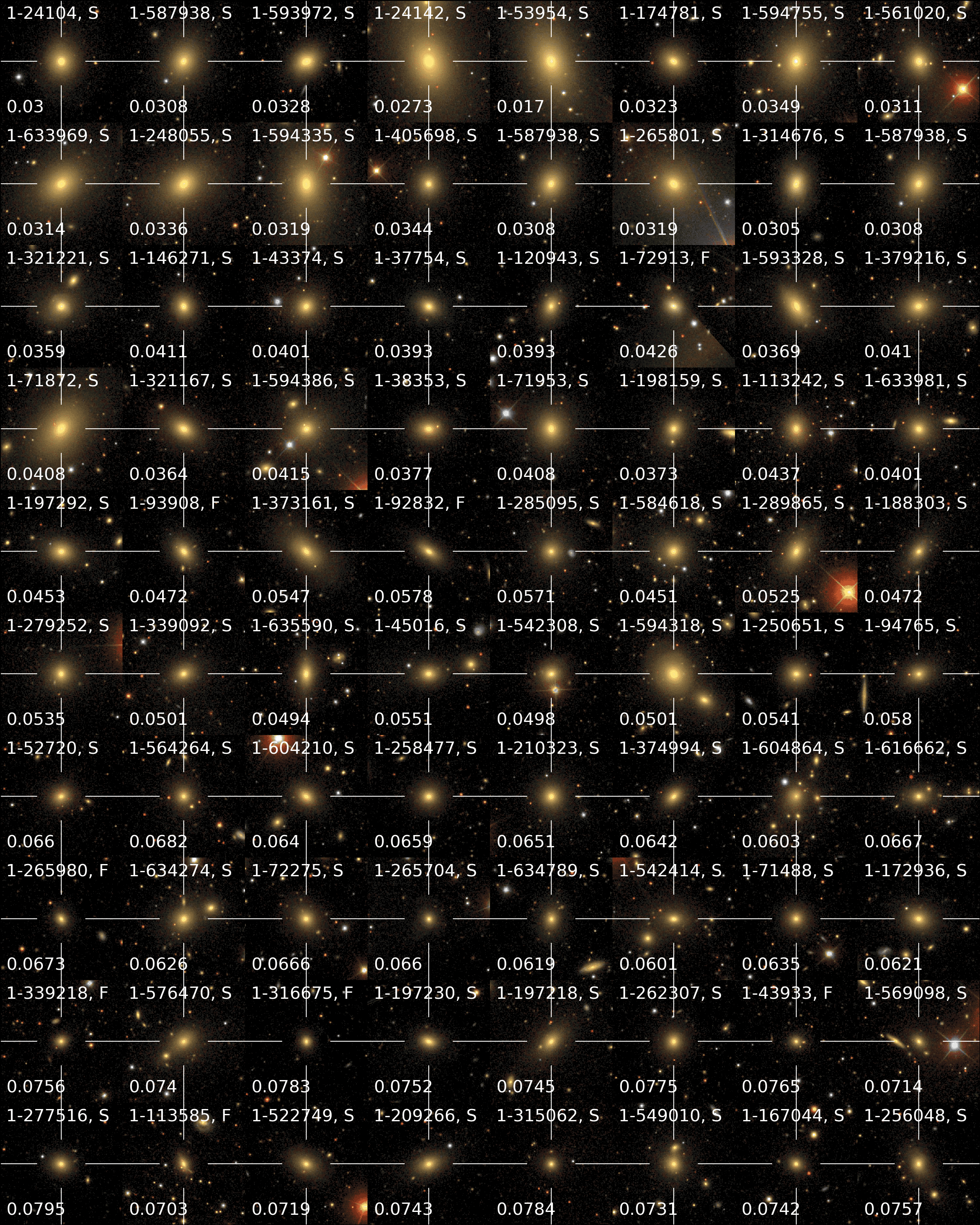}
\caption[Confirmed slow rotators]{\textbf{Confirmed slow rotators.} We show SDSS optical images for 80 randomly selected confirmed SRs. We select galaxies in five redshift bins with redshift increasing from the top downwards. The .jpg images have been adjusted to better highlight the faint outer envelopes that are characteristic of SRs. For each galaxy, we give the MaNGA-ID and the guess (upper left) and the redshift and stellar mass (lower left).}
\label{fig:Control1}
\end{figure*}
\begin{figure*}
\centering
\includegraphics[page=3,width=\textwidth]{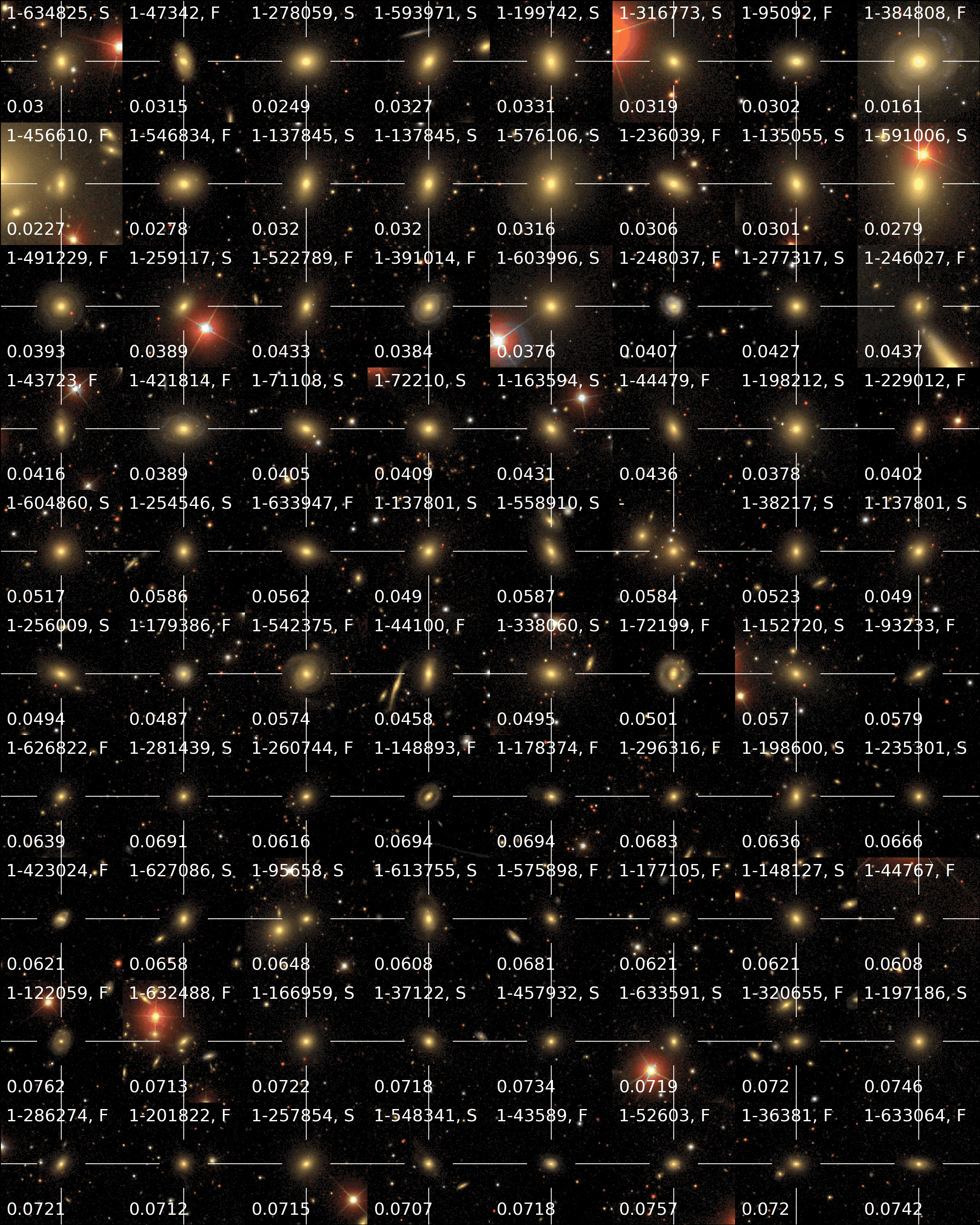}
\caption[Confirmed fast rotators]{\textbf{Confirmed fast rotators.} The same as \autoref{fig:Control1} except for ETGs which have been confirmed as fast rotators.}
\label{fig:Control2}
\end{figure*}
\begin{figure*}
\centering
\includegraphics[page=2,width=\textwidth]{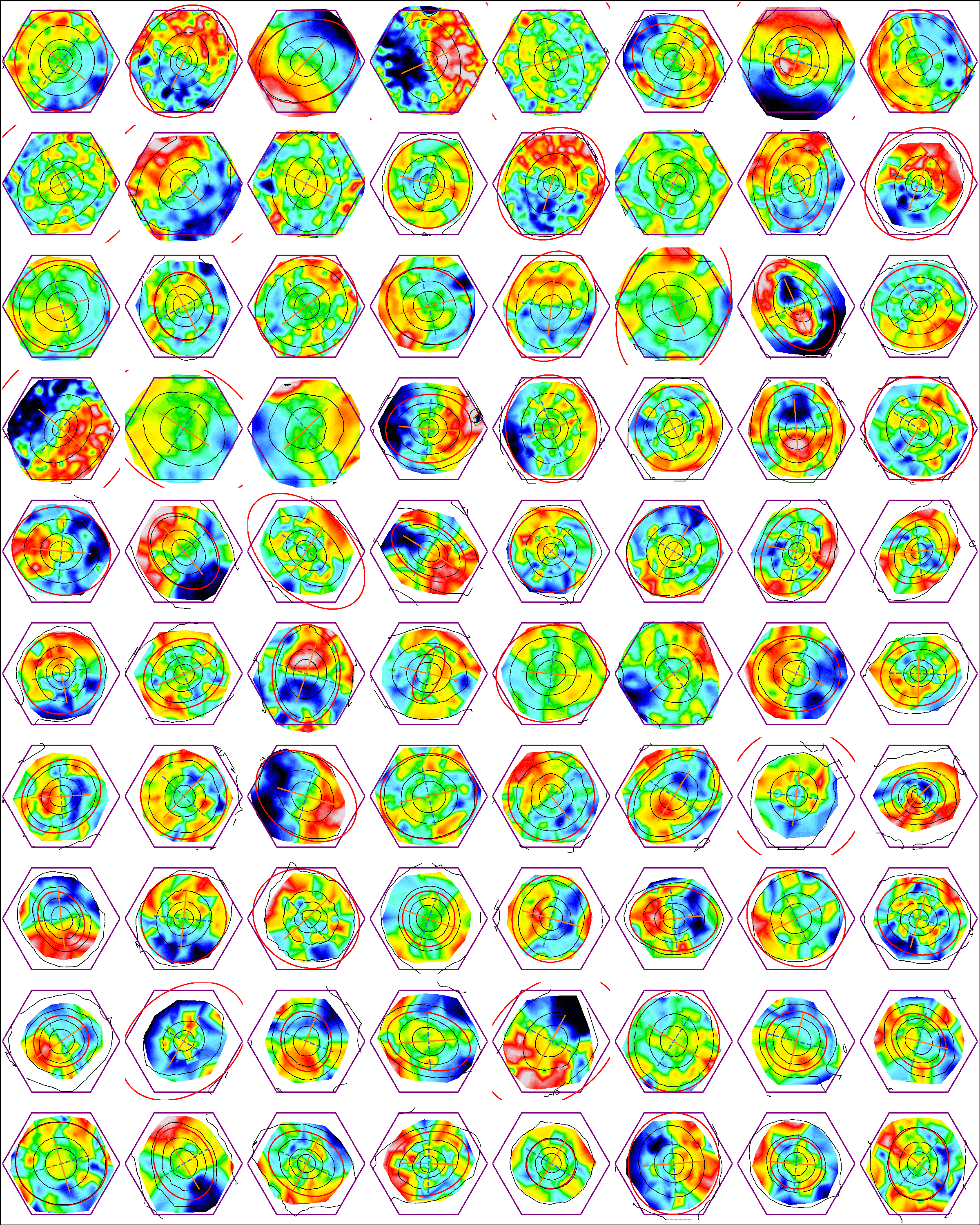}
\caption[Slow rotator kinematics]{\textbf{Slow rotator kinematics.} For each galaxy in \autoref{fig:Control1}, we plot the velocity map from the MaNGA stellar kinematics. Velocities are red and blue shifted according to their colour and green if they are at the mean velocity i.e. zero. The red ellipse indicates the half-light ellipse with an area equal to $\pi R_e^2$. In many cases, there is little sign of rotation. If the maximum speed is less than 30 km s$^{-1}$, then we fix the colourmap to range between -30 and + 30 km s$^{-1}$. }
\label{fig:Controlkine1}
\end{figure*}
\begin{figure*}
\centering
\includegraphics[page=4,width=\textwidth]{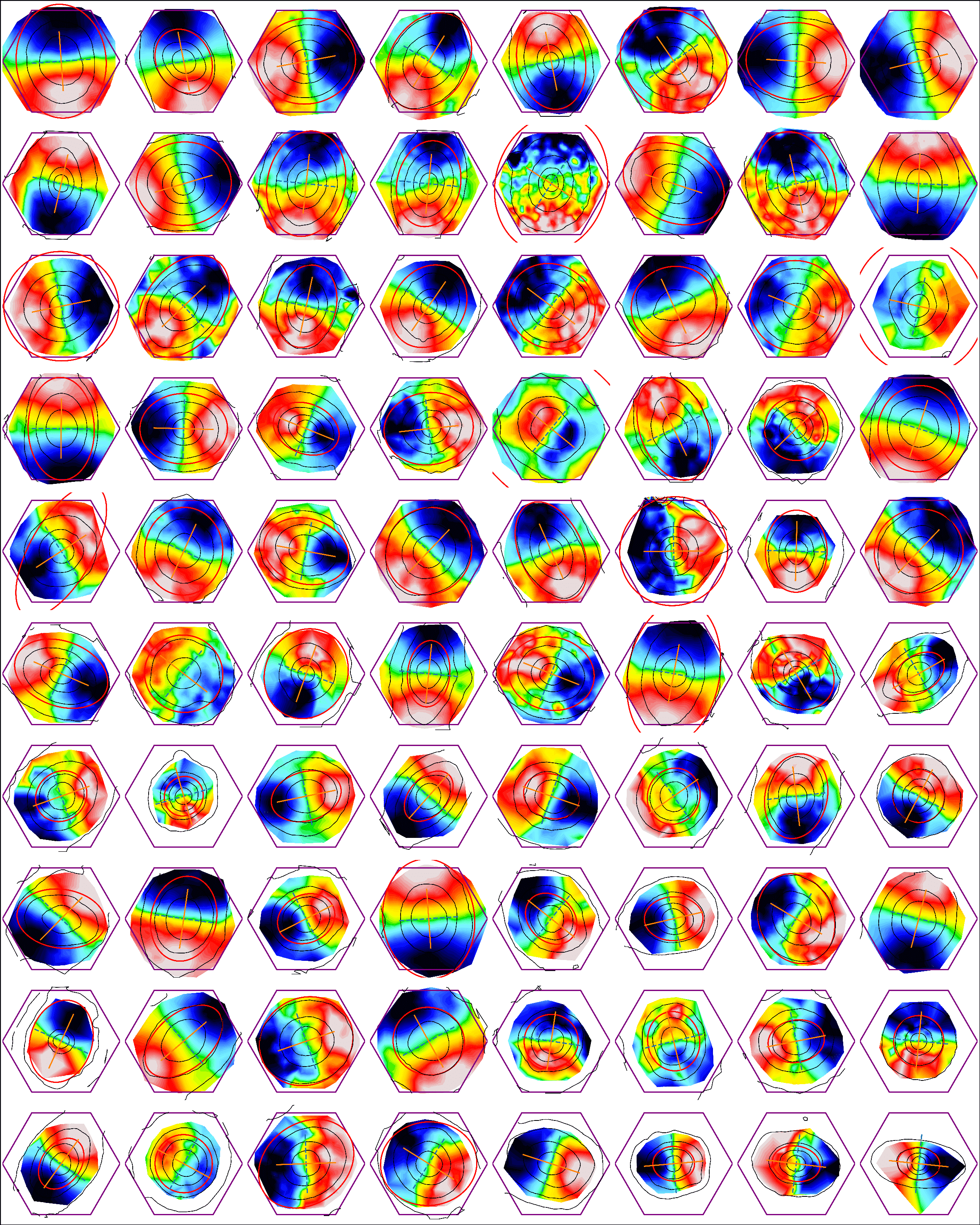}
\caption[Fast rotator kinematics]{\textbf{Fast rotator kinematics.} The same as \autoref{fig:Controlkine1} except for fast rotators. In almost all cases, the characteristic hourglass velocity field is clearly visible.}
\label{fig:Controlkine2}
\end{figure*}

Before we quantify our rate of success, we present a random selection of 160 MaNGA ETGs that are \textit{confirmed} SRs and FRs (see \autoref{fig:Control1}, \autoref{fig:Control2} for true-colour images and \autoref{fig:Controlkine1}, \autoref{fig:Controlkine2} for velocity maps). To aid our comparison, we split up the MaNGA ETGs into five redshift bins and randomly select 32 galaxies in each bin. For each galaxy, we give the MaNGA-ID, the redshift and our guess (S/F).
While it is obvious which of these galaxies are \textit{not} SRs, there are a substantial number of false positives (FRs incorrectly classified as SRs). From the images, it becomes clear why we decided to limit this classification to $z\leq0.08$: it simply becomes harder to pick out the defining morphological features pertaining to FRs and SRs when the redshift is large.

\subsection{Quantifying rate of success}
\label{sec:success}

\begin{figure}
\centering
\includegraphics[width=0.49\textwidth]{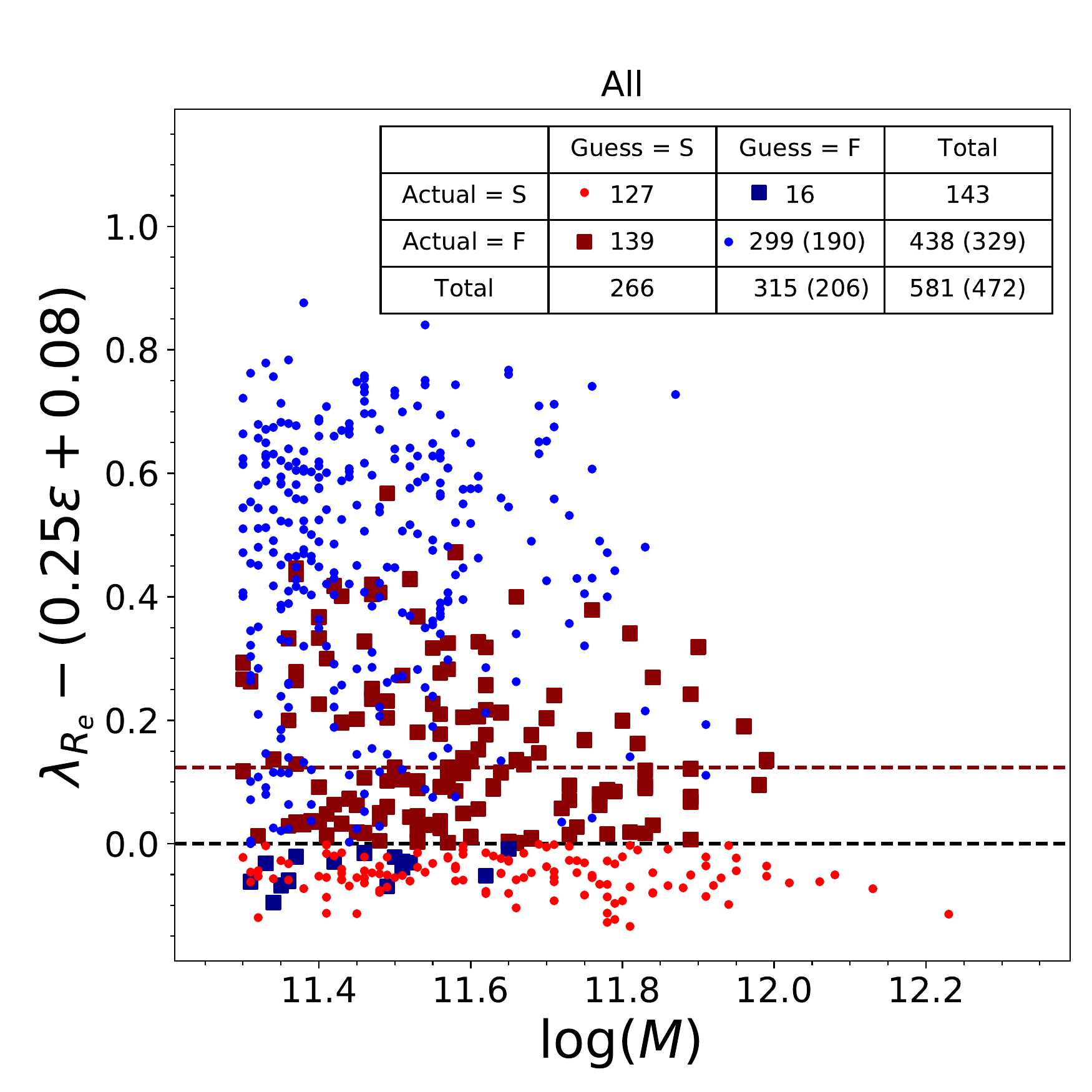}
	\caption{\textbf{Quantifying success rates.} We plot the vertical distance from the upper SR limit as a function of stellar mass for all MaNGA galaxies which satisfy the SR photometric criteria i.e. $\log(M) \geq 11.3$ and $\epsilon < 0.4$. The sample is split into four subsamples: true positives (red circles), true negatives (blue circles), false positives (dark red squares) and false negatives (dark blue squares). The confusion matrix is given in the upper right and details the number of galaxies in each subsample. Where two numbers are given, the number in brackets is for ETGs only, and the number outside the brackets includes late types. The black horizontal dashed line indicates the SR boundary and separates positives (below) from negatives (above). The dark red horizontal dashed line indicates the median vertical distance for the false positives.}
	\label{fig:True_false_mass_lam}
\end{figure}
\begin{figure}
\centering
	\includegraphics[width=0.49\textwidth]{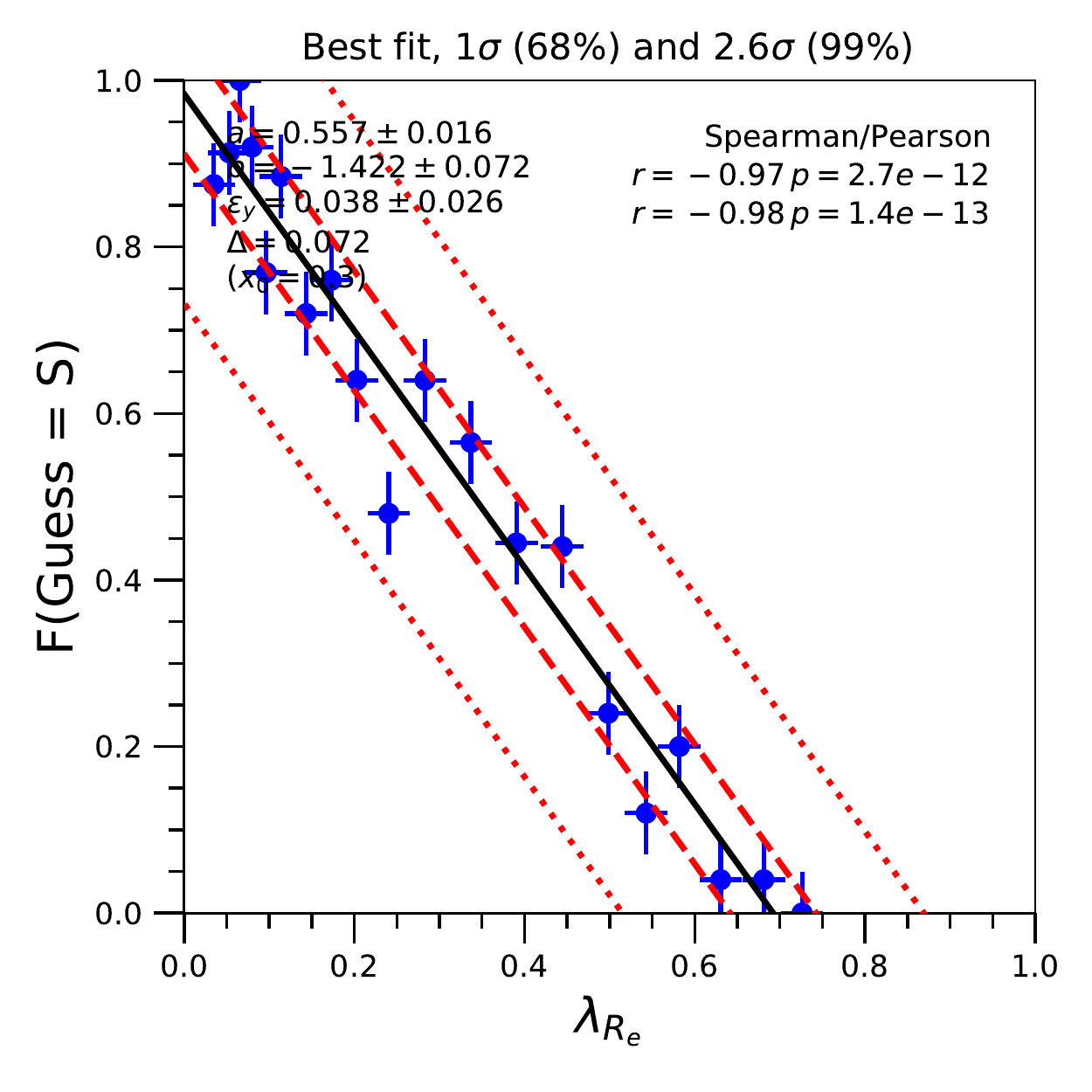}
	\caption{\textbf{Estimating $\lambda_{R_e}$ by eye.} We plot the fraction of galaxies classified visually as slow rotators regardless of whether they are confirmed as slow rotators or not. The parameters of the linear fit are shown.}
	\label{fig:frac_lambda}
\end{figure}

In \autoref{fig:True_false_mass_lam}, we plot $\lambda_{R_e} - (0.25\epsilon + 0.08)$, which is the vertical distance from the upper limit for SRs, as a function of stellar mass for MaNGA galaxies of all morphologies that satisfy the SR criteria (i.e. $\log(M) \geq 11.3$, $\epsilon<0.4$). All galaxies that lie above and below the dashed line are \textit{confirmed} FRs and SRs respectively. All blue points (dark squares and light circles) have been classified \textit{by eye} as FRs, and the red equivalent have been visually classified as SR \textit{candidates}. We also give the confusion matrix for the sample which contains our key result: Crucially, the number of false negatives is only 16 out of a total of 581 galaxies. The fact that this number is so small is pivotal for the accuracy of our angular momentum classification for two key reasons. Firstly, it means that 127/143 ($\sim89\%$) of \textit{confirmed} SRs are visually classified as such. Hence, we are confident that we miss only one in ten genuine massive SRs. Secondly, and more importantly, \textit{given a guess of FR}, we are correct for 299 out of 315 cases, giving us a success rate of $\sim95\%$. If we consider only ETGs, this fraction drops slightly to 190/206 ($\sim92\%$). This means that if we classify 100 galaxies as FRs, then only 5 will have incorrect classifications (and will actually be SRs). Furthermore, this \textit{only} applies to galaxies with $11.3 \leq \log(M) \lesssim 11.7$. Above this approximate upper limit in stellar mass, we are able to classify FRs with very high accuracy, and we also do not miss any genuine SRs.\par
On the contrary, of the 266 galaxies we classify as SRs, only 127 are actually SRs, which means that \textit{given a guess of SR}, we are only correct in $\sim48\%$ of cases. One reason for this is that all SRs have the morphological features described in \autoref{sec:visual_class}, but not all galaxies with those features are necessarily SRs. Another reason is that we take into account the stellar mass in making our classifications. The SR fraction is a strong function of stellar mass, and so the more massive a galaxy, the more likely we are to classify it as a SR if it has the required morphology. Reassuringly, 50\% of the false positives are within +0.125 of the upper limit for SRs ($\lambda_{R_e}=0.08+0.25\epsilon$; see \autoref{fig:True_false_mass_lam}) which, given that $\lambda_{R_e}$ has errors of about 0.05 in this $\lambda_{R_e}$ regime \citep{yan2016sdss}, means that we are able to classify galaxies that have a non-negligible chance of actually being in the SR regime. In fact, the fraction of false positives decreases linearly with increasing $\lambda_{R_e}$ with surprisingly small scatter (\autoref{fig:frac_lambda}), and very few galaxies with $\lambda_{R_e}>0.5$ are guessed as SRs. This implies that the eye is able to qualitatively estimate the velocity dispersion in ETGs with remarkable accuracy.\par
There is the caveat that the fractions quoted above are not for a volume weighted sample. However, because we are only considering massive galaxies, the volume weighting has a minimal effect. Nevertheless, we also give the volume weighted percentages. Our misclassification rate for SRs increases from $\sim11\%$ to $\sim26\%$ for the volume weighted sample because the SRs that are guessed as FRs are lower than $\log(M)\approx11.7$ in stellar mass and so they are relatively upweighted. Hence, we miss one in four in the volume weighted sample. Crucially, given a guess of FR, we are still correct in $\sim95\%$ of all cases ($\sim93\%$ for ETGs only) even for the volume weighted sample. Unfortunately, our true positive rate decreases from $\sim48\%$ to $\sim39\%$ after applying the volume weighting. As previously mentioned, the volume weighted sample is constructed artificially by duplicating galaxies, and of course no two galaxies in the Universe are identical. Hence, there is some uncertainty on the volume-weighted true positive rate that is difficult to quantify.\par
Finally, as an aside, we check to see if we have improved on the two-thirds misclassification rate for \textit{elliptical} galaxies as quoted in \cite{cappellari2011atlas3db}. Only two MaNGA galaxies were classified as SR but not as elliptical galaxies (i.e. S0s), and so the fraction is essentially the same as quoted above. Therefore, we have been able to improve on the misclassification rate for SRs given in \cite{cappellari2011atlas3db}.\par
As an aside, the two galaxies which were separately classified as S0s and SRs, 1-594069 and 1-156062, illustrate the difficultly in classifying galaxies with morphology. Both these galaxies were classified on separate occasions to be disky (S0) and spheroidal (SR) by MG. However, this only happened for $\sim1\%$ of MaNGA galaxies that were guessed to be SRs and so the classifications are in general very consistent.

\subsection{Example candidates}
\begin{figure*}
\centering
\includegraphics[page=1,width=\textwidth]{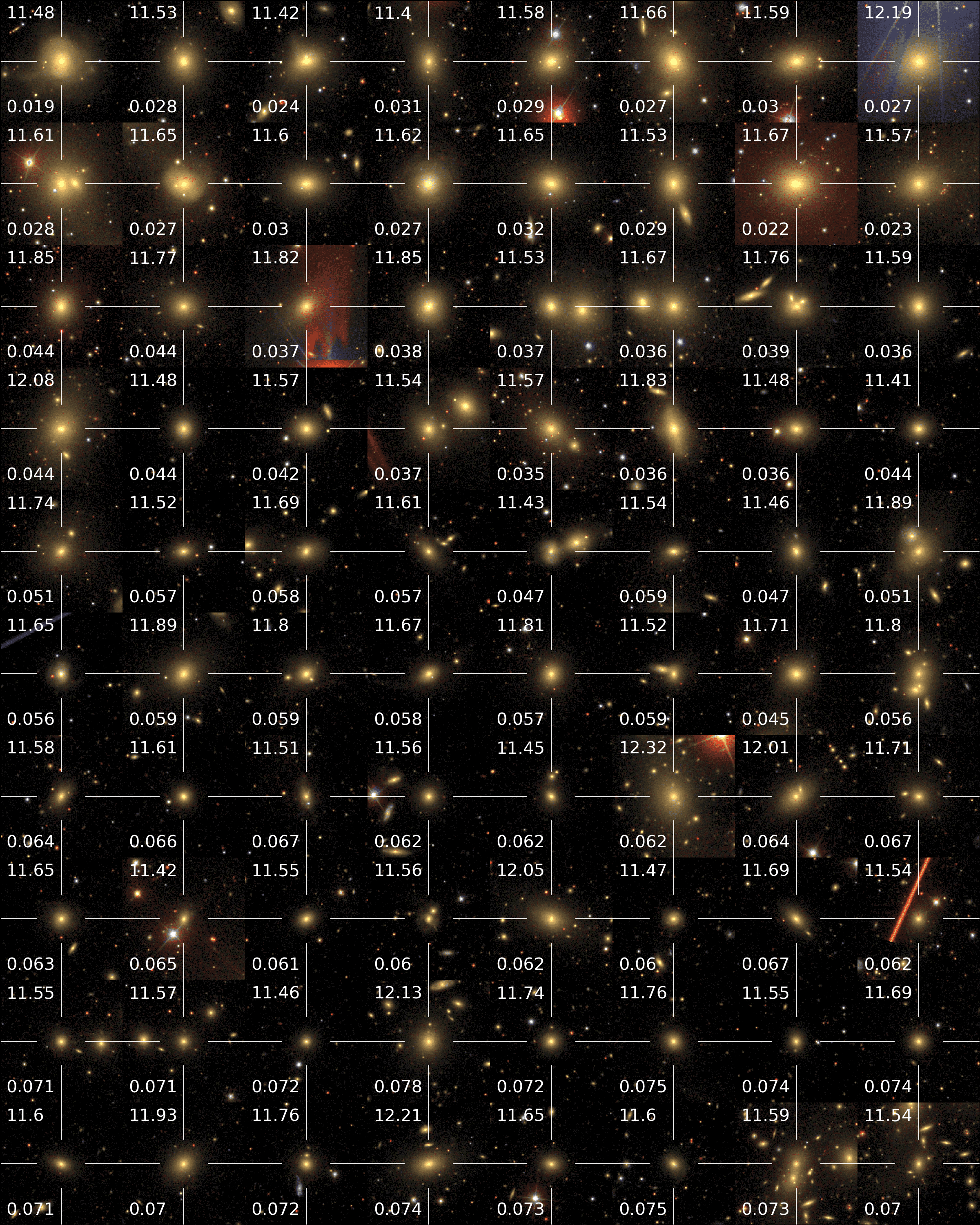}
\caption[Slow rotator candidates]{\textbf{Slow rotator candidates.} We show SR candidates using the same bins as \autoref{fig:Control1} etc. For each galaxy, we give the stellar mass (upper left) and the redshift (lower left) as we take these into account when assigning a classification. According to our analysis of the MaNGA sample, about half of these galaxies will be confirmed as FRs if we were able to obtain stellar kinematics (see text).}
\label{fig:Sample1}
\end{figure*}

\begin{figure*}
\centering
\includegraphics[page=2,width=\textwidth]{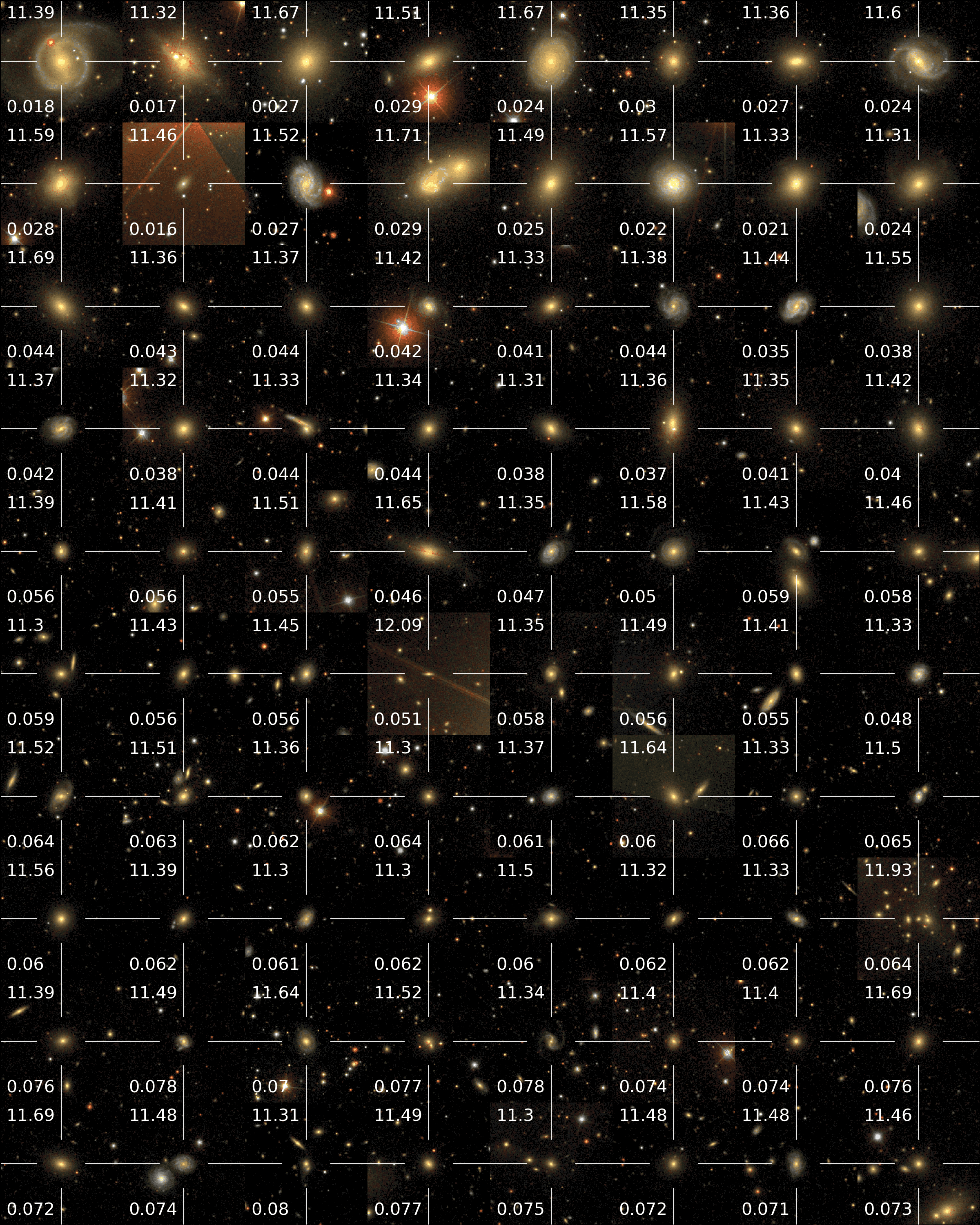}
\caption[Fast rotator candidates]{\textbf{Fast rotator candidates.} The same as \autoref{fig:Sample1} except for galaxies which we have classified as FRs. According to our analysis of the MaNGA sample, only about 6\% of these galaxies will be confirmed as SRs with observations of the stellar kinematics.}
\label{fig:Sample2}
\end{figure*}

In \autoref{fig:Sample1}, we present a random selection of the 262 galaxies we visually classify as SRs. All these galaxies share morphological characteristics with the SRs in \autoref{fig:Control1}, but one half will actually turn out to be FRs if/when stellar kinematics become available. \autoref{fig:Sample2} shows a random selection of the 203 FR candidates. The majority of these are unambiguous in their classification since they possess disk-like features. A handful appear to be ellipticals but are deemed to either be too elongated or likely to contain disks.\par



\subsection{Catalog statistics}
In \autoref{tab:sample_stats}, we summarise the number of galaxies in each sample. We only consider galaxies in unique groups that have been selected using the algorithm described in sec. 5 of \citetalias{graham2019atechnical}. This means that no MaNGA galaxy can appear more than once in our catalogue, but galaxies that are not in MaNGA can be duplicated.\par
\begin{table*}
\centering
\caption{Table containing the numbers of galaxies in the various samples used in this work, which are labelled in Column (1). Column (2) lists the criteria used to define the sample given in Column (1), and Column (3) gives the number of total number of galaxies in each sample. Columns (4) and (5) give the number of MaNGA galaxies of all morphologies and ETGs respectively. Column (6) lists the number of galaxies which are not in MaNGA that are contained in our catalogue. This column is split up into those with spectroscopic (Column (7)) and photometric (Column (8)) redshifts.The groups that make up each sample have been selected using our group selector algorithm (see \citetalias{graham2019atechnical}). As such, the MaNGA galaxies in each sample are unique, but of the galaxies which are not in MaNGA, there may be a small number of duplicates.}
\resizebox{\textwidth}{!}{%
\begin{tabular}{@{}cc|c|cc|ccc@{}}
\toprule
\multirow{2}{*}{\begin{tabular}[c]{@{}c@{}}Sample\\ \\(1)\end{tabular}} &
\multirow{2}{*}{\begin{tabular}[c]{@{}c@{}}Criteria\\ \\(2)\end{tabular}} &
\multirow{2}{*}{\begin{tabular}[c]{@{}c@{}}Total\\ \\(3)\end{tabular}} &
\multicolumn{2}{c|}{MaNGA} &
\multicolumn{3}{c}{Non-MaNGA} \\ \midrule
 & & & 
\begin{tabular}[c]{@{}c@{}}Total\\(4)\end{tabular} &
\begin{tabular}[c]{@{}c@{}}ETGs\\(5)\end{tabular} &
\begin{tabular}[c]{@{}c@{}}Total\\(6)\end{tabular} &
\begin{tabular}[c]{@{}c@{}}Spectroscopic $z$\\(7)\end{tabular} &
\begin{tabular}[c]{@{}c@{}}Photometric $z$\\(8)\end{tabular} \\ \midrule
A & All & 17300 & 4460 & 3168 & 12840 & 8177 & 4663\\
B & A \& $z\leq0.08$ & 14093 & 3891 & 2761 & 10202 & 6683 & 3519\\
C & B \& $\log(M_*)\geq11.3$ & 1594 & 812 & 588 & 782 & 749 & 33\\
D & C \& $\epsilon<0.4$ & 1169 & 616 & 479 & 553 & 529 & 24\\
E & D \& ROT = S & 345 & 174 & 154 & 171 & 171 & 0\\
\bottomrule
\label{tab:sample_stats}
\end{tabular}}
\end{table*}

Sample A consists of all groups in the MaNGA redshift range ($0.01 \leq z \leq 0.15$). There are 4597 unique MaNGA galaxies in MPL-7, but only 4430 in our catalogue. This is because of our random group selection process that selects representative groups from duplicate sets (see sec. 5 of \citetalias{graham2019atechnical}). As discussed in \autoref{sec:maximum_z}, we limit our volume-limited sample with complete angular momentum classifications to $z\leq0.08$, which we refer to in \autoref{tab:sample_stats} as Sample B. By introducing a lower maximum redshift, we only reduce our sample size by $\sim18.5\%$ and so we do not significantly compromise on number statistics.\par
Next, we define Sample C to be a subsample contained within Sample B that itself only contains galaxies more massive than $M_{\rm{crit}}$. About 20.9\% of MaNGA galaxies that are in Sample B also lie in Sample C, whereas only $\sim7.7\%$ of non-MaNGA galaxies satisfy the mass threshold. This is not surprising since MaNGA has a flat selection in stellar mass, while massive galaxies are much rarer in the general population (non-MaNGA). Above $M_{\rm{crit}}$, the redshift accuracy is much higher than for the whole sample, where only $\sim2.1\%$ of galaxies above the mass limit have photometric redshifts.\par
We also define the sample which we classify using our visual method (Sample D). The number of MaNGA galaxies is given as 616 which is slightly higher than the MaNGA sample used in \autoref{sec:visual_class}. This number includes mergers and galaxies with flagged kinematics. We do not use $\lambda_{R_e}$ or $\epsilon$ for merging galaxies to quantify the angular momentum because one or both of these can be inaccurate for a merger. Hence, we classify the merger galaxy centred in the IFU  as a fast or slow rotator using the same technique as described in \autoref{sec:visual_class}. Because we do not know the true angular momentum for merging galaxies, we do not include them in our training set. However, we do not exclude them from our catalogue on that basis. It is shown that in total, we classify 553 candidates using our visual method.\par
Finally, we define Sample E which only contains confirmed and candidate SRs. Of these galaxies, $\sim49.6\%$ are candidates of which $\sim52\%$ will be misclassified. In total, about a quarter of SRs in our catalogue will be misclassified FRs. None of the galaxies in Sample E have photometric redshifts so their proximity to MaNGA galaxies is robust.
\section{Conclusions}
\label{sec:conclusions1}
In this paper, we have produced a volume-limited sample of about 14000 galaxies with complete estimates for stellar mass, ellipticity and angular momentum. This is the largest sample of its kind available in the literature and has been checked for quality and reliability. In \autoref{sec:photometry}, we introduced four new criteria for selecting a sample of galaxies with clean photometry that are independent from the flags provided by the SDSS imaging pipeline. The first of these criteria are limits in five colours that effectively clip outliers with unusual colours (see \autoref{fig:color-color}). We find that most objects that are clipped are faint and can have inaccurate photometric measurements due to fitting errors. However, a small fraction of excluded objects are in galaxy clusters. We also examine the nature of very bright objects with small Petrosian radii using infrared colours from the 2MASS PSC and find that they are in fact foreground stars (\autoref{fig:2MASS_colors}). The stars are extended and so were missed by the SDSS pipeline, which is optimised to classify point sources as stars (\autoref{fig:2MASS_stars}). We suggest that for the bright end ($m_r<16$), a minimum value of $R_{\rm{r,Petro}}=1.5$ arcsec should be enforced (\autoref{fig:re_cuts}). Of course, these objects could also be removed on the basis of their $J-K_S$ colour, but having equivalent criteria defined using just the SDSS photometric quantities negates the need to consult the 2MASS PSC. For the range $16<m_r\leq19$, a maximum error in $z_{\rm{phot}}$ should be limited to 0.075 (\autoref{fig:high_zerr}).\par
We took the photometric redshifts that were estimated by \cite{beck2017redshift} for all objects that were classified as galaxies. We adopted a selection criteria that only accepts redshifts with a certain degree of accuracy and reliability. In \autoref{fig:frac_z_zphot}, we showed that galaxies with photometric redshifts that do not satisfy these criteria only constitute a significant fraction fainter than $m_r \approx 20.2$ mag. We confirm that our new criteria are independent of the $r$-band flags provided by the SDSS pipeline and so cannot be recreated using only these flags (\autoref{fig:sdss_flags}). One particular flag, \texttt{NOPETRO\_BIG}, is assigned to objects where the signal-to-noise ratio is small and hence the fit is extended out to unphysically large radii. We find that our colour criteria almost eliminates the objects with very large radii, but ignores the objects that have the flag but are \textit{intrinsically} faint and actually have small $R_{\rm{r,Petro}}$ (\autoref{fig:nopetro_big}). Hence, by applying our criteria along with the criteria recommended for clean photometry (see \autoref{sec:clean}), a clean photometric galaxy sample can be obtained.\par
We estimate the stellar mass for all galaxies in the combined NSA/photometric catalogue by fitting $M_r$ to the dynamical masses from MaNGA (\autoref{fig:r_band_stellar_mass}). In \autoref{fig:fluxes}, we show that our combined sample is complete down to a stellar mass of about $7.4 \times 10^{9} \textrm{ M}_{\odot}$, up until the maximum redshift of MaNGA. To improve the selection and remove potential interlopers with large photometric redshifts, we make the important assumption that all neighbours of a MaNGA galaxy should have an absolute luminosity that satisfies the minimum stellar mass ($M_r\leq-18$), assuming all neighbours are at the redshift of the MaNGA galaxy. By making this assumption, we are able to exclude a large fraction of interlopers that overlap in redshift but would not satisfy our luminosity threshold if they are, in fact, bound to the halo containing the MaNGA galaxy (\autoref{fig:frac_mag_cut}).\par
From each set of surviving galaxies that surrounds each MaNGA galaxy, we find the group enclosing the galaxy using the new algorithm \texttt{TD-ENCLOSER}, which is described in detail in \citetalias{graham2019atechnical} and is optimised for obtaining the local galaxy environment. Because of the decreasing accuracy of the groups with increasing redshift due to the higher fraction of photometric galaxies, we limit our final sample to groups below $z=0.08$ (see \autoref{sec:volume-limited}). For all galaxies in the group catalogue which are not observed with MaNGA, we assign angular momentum classifications using a novel visual classification method (\autoref{sec:visual_class}). We test our method on a sample of 581 galaxies that satisfy the mass and shape criteria for genuine SRs (see \autoref{fig:Control1}, \autoref{fig:Control2} for true-colour images and \autoref{fig:Controlkine1}, \autoref{fig:Controlkine2} for velocity maps of a random selection). We find that we can determine which galaxies are fast rotators with an accuracy of $\sim94\%$ (\autoref{fig:True_false_mass_lam}). However, we are only correct in about half of all cases where we guess a galaxy to be a SR. This is a slight improvement on the SR positive identification rate of 34\% for ETGs found in ATLAS$^{\rm{3D}}$. A surprising result is that the eye is qualitatively able to estimate the velocity dispersion in galaxies, as the majority of galaxies we incorrectly identify as SRs are close to the SR boundary (see \autoref{fig:frac_lambda}). Crucially, we always correctly identify SRs if they are more massive than about $10^{11.7} \textrm{ M}_{\odot}$. The galaxies which we classify as SRs should be considered as SR \textit{candidates} and would make good targets for future observations with IFS (see \autoref{fig:Sample1} and \autoref{fig:Sample2} for candidate SRs and FRs respectively).\par
In \autoref{tab:Group_table}, \autoref{tab:MaNGA_table} and \autoref{tab:Not_MaNGA_table}, we tabulate a small portion of our group catalogue. In \citetalias{graham2019cenvironment}, we use this catalogue to conduct the largest study of galaxy kinematics and the kT-$\Sigma$ relation to date. The high level of accuracy overall for our angular momentum classifications, combined with its large scope, makes our catalogue the benchmark for future studies of this kind, and has the potential to provide powerful constraints on galaxy evolution.
\begin{table*}
\centering
\caption{Table containing the results of our visual classification test. All the quantities required to plot \autoref{fig:True_false_mass_lam} and calculate the confusion matrix are given. Column (1) is the MaNGA-ID, Column (2) is the confirmed classification from the $\lambda_{R_e}$ diagram, Column (3) is the visual classification, and Columns (4) and (5) are the $y$ and $x$ axes of \autoref{fig:True_false_mass_lam} respectively. Only the first 10 rows are shown.}
\resizebox{0.700\textwidth}{!}{%
\begin{tabular}{@{}ccccc@{}}
\toprule
\begin{tabular}[c]{@{}c@{}}MaNGA ID\\\\(1)\end{tabular} &
\begin{tabular}[c]{@{}c@{}}Confirmed\\Classification\\(2)\end{tabular} &
\begin{tabular}[c]{@{}c@{}}Visual\\Classification\\(3)\end{tabular} &
\begin{tabular}[c]{@{}c@{}}$\lambda_{R_e}-(0.25\epsilon + 0.08)$\\\\(4)\end{tabular} &
\begin{tabular}[c]{@{}c@{}}Stellar Mass\\$\log(\rm{M})$\\(5)\end{tabular} \\ \midrule
1-178828	&	F	&	F	&	0.369	&	11.52	\\
1-384808	&	F	&	F	&	0.291	&	11.42	\\
1-53954	&	S	&	S	&	-0.1198	&	11.32	\\
1-456984	&	F	&	F	&	0.088	&	11.54	\\
1-593929	&	S	&	S	&	-0.049	&	11.48	\\
1-604022	&	F	&	F	&	0.657	&	11.32	\\
1-456610	&	F	&	F	&	0.3848	&	11.47	\\
1-256293	&	F	&	F	&	0.5875	&	11.33	\\
1-278059	&	F	&	S	&	0.2932	&	11.3	\\
1-392067	&	F	&	F	&	0.512	&	11.33	\\
\bottomrule
\label{tab:MaNGA_guess}
\end{tabular}}
\end{table*}


\section*{Acknowledgements}
Funding for the Sloan Digital Sky Survey IV has been provided by the Alfred P. Sloan Foundation, the U.S. Department of Energy Office of Science, and the Participating Institutions. SDSS acknowledges support and resources from the Center for High-Performance Computing at the University of Utah. The SDSS website is www.sdss.org.\par
SDSS is managed by the Astrophysical Research Consortium for the Participating Institutions of the SDSS Collaboration including the Brazilian Participation Group, the Carnegie Institution for Science, Carnegie Mellon University, the Chilean Participation Group, the French Participation Group, Harvard-Smithsonian Center for Astrophysics, Instituto de Astrofísica de Canarias, The Johns Hopkins University, Kavli Institute for the Physics and Mathematics of the Universe (IPMU) / University of Tokyo, Lawrence Berkeley National Laboratory, Leibniz Institut für Astrophysik Potsdam (AIP), Max-Planck-Institut für Astronomie (MPIA Heidelberg), Max-Planck-Institut für Astrophysik (MPA Garching), Max-Planck-Institut für Extraterrestrische Physik (MPE), National Astronomical Observatories of China, New Mexico State University, New York University, University of Notre Dame, Observatório Nacional / MCTI, The Ohio State University, Pennsylvania State University, Shanghai Astronomical Observatory, United Kingdom Participation Group, Universidad Nacional Autónoma de México, University of Arizona, University of Colorado Boulder, University of Oxford, University of Portsmouth, University of Utah, University of Virginia, University of Washington, University of Wisconsin, Vanderbilt University, and Yale University.\par
This publication makes use of data products from the Two Micron All Sky Survey, which is a joint project of the University of Massachusetts and the Infrared Processing and Analysis Center/California Institute of Technology, funded by the National Aeronautics and Space Administration and the National Science Foundation.





\bibliographystyle{mnras}
\bibliography{MasterBibliography} 



\appendix
\section{Group and galaxy catalogues}
\label{sec:appendix}
\begin{table*}
\centering
\caption{Table containing the group characteristics. Column (1) gives the group number and Column (2) gives the group richness. Column (3) gives the group radius equal to the 90th percentile of $D_{\rm{cen}}$. Columns (4) and (5) give the coordinates of the central galaxy of the group and Column (6) gives the redshift of the host MaNGA galaxy (see \citetalias{graham2019atechnical}). Only the first 10 rows are shown.}
\resizebox{0.600\textwidth}{!}{%
\begin{tabular}{@{}cccccc@{}}
\toprule
\begin{tabular}[c]{@{}c@{}}Group\\Number\\(1)\end{tabular} &
\begin{tabular}[c]{@{}c@{}}Group\\Richness $\mathcal{N}$\\(2)\end{tabular} &
\begin{tabular}[c]{@{}c@{}}90th \%\\(Mpc)\\(3)\end{tabular} &
\begin{tabular}[c]{@{}c@{}}RA\\(\degree)\\(4)\end{tabular} &
\begin{tabular}[c]{@{}c@{}}DEC\\(\degree)\\(5)\end{tabular} &
\begin{tabular}[c]{@{}c@{}}Redshift of\\host galaxy\\(6)\end{tabular} \\ \midrule
1	&	1	&	0.0	&	230.50746	&	43.53234	&	0.0205	\\
2	&	1	&	0.0	&	231.3607	&	43.1699	&	0.0575	\\
3	&	1	&	0.0	&	229.52558	&	42.74584	&	0.0403	\\
4	&	1	&	0.0	&	230.15302	&	41.96044	&	0.0645	\\
5	&	3	&	0.347	&	231.20531	&	42.17256	&	0.0188	\\
6	&	1	&	0.0	&	230.57438	&	42.28702	&	0.0179	\\
7	&	1	&	0.0	&	230.59834	&	43.36778	&	0.1105	\\
8	&	1	&	0.0	&	232.16734	&	43.02268	&	0.0286	\\
9	&	1	&	0.0	&	231.47876	&	41.90977	&	0.0227	\\
10	&	1	&	0.0	&	231.57732	&	41.29463	&	0.0309	\\
\bottomrule
\label{tab:Group_table}
\end{tabular}}
\end{table*}

\begin{sidewaystable*}
\centering
\caption{Table containing the relevant information for the MaNGA galaxies in our sample. Descriptions for columns 1-16 are found in the caption of \cite{graham2018angular}. Column 17 is the final group number that the MaNGA galaxy appears in and can be used to match up with \autoref{tab:Group_table}. Columns 18, 19, 20, 21 and 22 are $\log(\Sigma_3)$, $\log(\Sigma_3^{\rm{rel}})$, $\log(\Sigma_{10})$, $D_{\rm{cen}}$, $D_{\rm{cen}}^{\rm{norm}}$ respectively. Only the first 30 rows are shown.}
\resizebox{1.000\textwidth}{!}{%
\begin{tabular}{@{}cccccccccccccccccccccc@{}}
\toprule
\begin{tabular}[c]{@{}c@{}}MaNGA-ID\\ \\(1)\end{tabular} &
\begin{tabular}[c]{@{}c@{}}Clean\\Sample?\\(2)\end{tabular} &
\begin{tabular}[c]{@{}c@{}}Hubble\\Group\\(3)\end{tabular} &
\begin{tabular}[c]{@{}c@{}}Kinematic\\Classification\\(4)\end{tabular} &
\begin{tabular}[c]{@{}c@{}}$\textrm{log}(R_e^{\rm{circ}})$\\(\arcsec)\\(5)\end{tabular} &
\begin{tabular}[c]{@{}c@{}}$\textrm{log}(R_e^{\rm{maj}})$\\(kpc)\\(6)\end{tabular} &
\begin{tabular}[c]{@{}c@{}}PSF\\(\arcsec)\\(7)\end{tabular} &
\begin{tabular}[c]{@{}c@{}}$\lambda_{R_e}^{\rm{true}}$\\ \\(8)\end{tabular} &
\begin{tabular}[c]{@{}c@{}}$\epsilon$\\ \\(9)\end{tabular} &
\begin{tabular}[c]{@{}c@{}}$\Psi_{\rm{phot}}$\\(\degree)\\(10)\end{tabular} &
\begin{tabular}[c]{@{}c@{}}$\Psi_{\rm{kin}}$\\(\degree)\\(11)\end{tabular} &
\begin{tabular}[c]{@{}c@{}}Error in $\Psi_{\rm{kin}}$\\(\degree)\\(12)\end{tabular} &
\begin{tabular}[c]{@{}c@{}}$\textrm{log}(M_\ast)$\\ \\(13)\end{tabular} &
\begin{tabular}[c]{@{}c@{}}$\textrm{log}(\sigma_e)$\\(km s$^{-1}$)\\(14)\end{tabular} &
\begin{tabular}[c]{@{}c@{}}$z$\\ \\(15)\end{tabular} &
\begin{tabular}[c]{@{}c@{}}$n$\\ \\(16)\end{tabular} &
\begin{tabular}[c]{@{}c@{}}Group\\Number\\(17)\end{tabular} &
\begin{tabular}[c]{@{}c@{}}$\log(\Sigma_3)$\\$Mpc^{-2}$\\(18)\end{tabular} &
\begin{tabular}[c]{@{}c@{}}$\log(\Sigma_3^{rel})$\\$Mpc^{-2}$\\(19)\end{tabular} &
\begin{tabular}[c]{@{}c@{}}$\log(\Sigma_{10})$\\$Mpc^{-2}$\\(20)\end{tabular} &
\begin{tabular}[c]{@{}c@{}}$D_{\rm{cen}}$\\$Mpc$\\(21)\end{tabular} &
\begin{tabular}[c]{@{}c@{}}$D_{\rm{cen}}^{\rm{norm}}$\\\\(22)\end{tabular} \\ \midrule
12-98126	&	Y	&	E	&	R	&	0.692	&	0.4	&	2.6	&	0.731	&	0.337	&	138.5	&	152.0	&	6.0	&	10.36	&	1.841	&	0.0205	&	4.4	&	1	&	-0.984	&	0.0	&	-0.854	&	0.0	&	0.0	\\
12-84674	&	N	&	U	&	R	&	0.939	&	0.996	&	2.5	&	0.77	&	0.048	&	100.6	&	110.5	&	8.2	&	11.1	&	1.799	&	0.0575	&	1.7	&	2	&	-0.23	&	0.0	&	-0.275	&	0.0	&	0.0	\\
12-193481	&	N	&	M/CP	&	R	&	0.775	&	0.739	&	2.6	&	0.425	&	0.25	&	24.9	&	165.5	&	12.5	&	11.33	&	2.111	&	0.0403	&	2.2	&	3	&	-0.755	&	0.0	&	-0.655	&	0.0	&	0.0	\\
12-84731	&	Y	&	S0	&	R	&	1.105	&	0.924	&	2.5	&	0.756	&	0.668	&	114.6	&	128.0	&	3.5	&	10.41	&	1.882	&	0.0187	&	0.9	&	4	&	0.124	&	0.0	&	0.139	&	0.0	&	0.0	\\
12-84627	&	Y	&	S0	&	R	&	0.806	&	1.09	&	2.6	&	0.88	&	0.586	&	40.5	&	42.0	&	3.8	&	11.2	&	2.108	&	0.0645	&	1.0	&	5	&	0.091	&	-0.22	&	-0.004	&	0.0	&	0.0	\\
1-322159	&	N	&	S0	&	R	&	0.649	&	0.29	&	2.6	&	0.344	&	0.235	&	77.9	&	58.0	&	15.0	&	9.78	&	1.897	&	0.0188	&	2.2	&	-999	&	-999.0	&	-999.0	&	-999.0	&	-999.0	&	-999.0	\\
1-199287	&	N	&	S0	&	R	&	0.551	&	0.307	&	2.6	&	0.697	&	0.547	&	10.5	&	6.5	&	10.8	&	10.01	&	1.624	&	0.0189	&	2.8	&	-999	&	-999.0	&	-999.0	&	-999.0	&	-999.0	&	-999.0	\\
1-322172	&	Y	&	I	&	2S	&	0.682	&	0.371	&	2.6	&	0.201	&	0.444	&	134.7	&	121.5	&	89.8	&	9.45	&	1.63	&	0.0179	&	1.3	&	-999	&	-999.0	&	-999.0	&	-999.0	&	-999.0	&	-999.0	\\
12-84670	&	Y	&	E	&	NR	&	0.544	&	0.855	&	2.5	&	0.024	&	0.032	&	111.7	&	68.5	&	89.8	&	11.84	&	2.314	&	0.1105	&	6.0	&	8	&	-0.49	&	0.0	&	-0.706	&	0.0	&	0.0	\\
12-180451	&	N	&	E	&	F	&	0.342	&	0.13	&	2.6	&	-999.0	&	0.121	&	143.5	&	-999.0	&	-999.0	&	9.3	&	-999.0	&	0.0286	&	0.9	&	9	&	-1.107	&	0.0	&	-0.806	&	0.0	&	0.0	\\
12-84617	&	N	&	E	&	R	&	0.529	&	0.31	&	2.6	&	0.169	&	0.423	&	65.6	&	11.0	&	43.0	&	9.49	&	1.916	&	0.0227	&	1.6	&	10	&	-0.892	&	0.0	&	-1.076	&	0.0	&	0.0	\\
12-84726	&	Y	&	S0	&	R	&	0.641	&	0.515	&	2.7	&	0.567	&	0.315	&	96.7	&	126.5	&	27.8	&	9.53	&	1.772	&	0.0309	&	1.1	&	11	&	-0.425	&	0.0	&	-0.645	&	0.0	&	0.0	\\
12-180432	&	Y	&	S	&	R	&	0.83	&	0.678	&	2.6	&	0.649	&	0.364	&	128.2	&	137.0	&	3.5	&	10.98	&	1.992	&	0.028	&	1.5	&	12	&	-0.189	&	0.0	&	0.016	&	0.0	&	0.0	\\
12-84665	&	Y	&	S0	&	R	&	0.772	&	0.516	&	2.6	&	0.481	&	0.548	&	15.3	&	28.5	&	3.8	&	9.98	&	1.925	&	0.0183	&	0.8	&	13	&	0.136	&	-0.111	&	0.293	&	0.0	&	0.151	\\
12-84677	&	Y	&	E	&	R	&	0.79	&	1.006	&	2.6	&	0.701	&	0.25	&	123.6	&	106.5	&	3.0	&	11.71	&	2.185	&	0.0751	&	3.1	&	14	&	-0.762	&	0.0	&	-0.749	&	0.0	&	0.0	\\
12-84660	&	Y	&	S	&	R	&	0.834	&	0.831	&	2.6	&	0.711	&	0.348	&	67.1	&	73.0	&	1.8	&	10.92	&	1.785	&	0.0405	&	2.4	&	15	&	-0.475	&	0.0	&	-0.435	&	0.0	&	0.0	\\
12-84679	&	Y	&	E	&	R	&	0.766	&	1.082	&	2.5	&	0.156	&	0.32	&	50.6	&	52.5	&	8.0	&	11.83	&	2.439	&	0.0916	&	6.0	&	16	&	0.17	&	0.0	&	0.436	&	0.0	&	0.0	\\
12-192120	&	Y	&	S0	&	R	&	0.747	&	0.866	&	2.3	&	0.522	&	0.193	&	120.8	&	143.0	&	6.2	&	11.25	&	2.149	&	0.0612	&	1.2	&	17	&	-0.624	&	0.0	&	-0.421	&	0.0	&	0.0	\\
12-129612	&	Y	&	S	&	R	&	0.879	&	0.958	&	2.3	&	0.899	&	0.775	&	13.6	&	13.5	&	1.5	&	11.05	&	2.071	&	0.0284	&	1.8	&	18	&	0.42	&	-0.327	&	-0.004	&	0.1	&	0.302	\\
1-458515	&	N	&	S0	&	R	&	1.007	&	0.944	&	2.3	&	0.922	&	0.522	&	15.1	&	15.0	&	2.0	&	10.88	&	2.012	&	0.0299	&	0.8	&	-999	&	-999.0	&	-999.0	&	-999.0	&	-999.0	&	-999.0	\\
1-458396	&	Y	&	S	&	R	&	1.057	&	0.938	&	2.3	&	0.902	&	0.418	&	151.3	&	171.5	&	1.5	&	11.17	&	2.061	&	0.0289	&	1.3	&	-999	&	-999.0	&	-999.0	&	-999.0	&	-999.0	&	-999.0	\\
12-129629	&	Y	&	S0	&	R	&	0.882	&	1.108	&	2.3	&	0.262	&	0.263	&	56.0	&	119.0	&	28.2	&	11.1	&	1.928	&	0.0762	&	4.7	&	20	&	-0.687	&	0.0	&	-0.767	&	0.0	&	0.0	\\
12-110746	&	Y	&	E	&	R	&	0.358	&	0.161	&	2.3	&	0.219	&	0.174	&	16.4	&	9.5	&	20.0	&	10.31	&	2.065	&	0.0288	&	4.9	&	21	&	0.628	&	-0.084	&	0.549	&	0.1	&	0.233	\\
12-129608	&	Y	&	S0	&	R	&	0.394	&	0.34	&	2.3	&	0.881	&	0.481	&	14.2	&	23.5	&	10.0	&	10.06	&	1.772	&	0.0318	&	2.2	&	22	&	-0.339	&	0.0	&	-0.234	&	0.0	&	0.0	\\
12-129446	&	Y	&	S0	&	R	&	0.77	&	0.911	&	2.3	&	0.754	&	0.138	&	145.2	&	156.5	&	3.0	&	11.5	&	2.076	&	0.067	&	2.0	&	23	&	0.303	&	0.0	&	-0.136	&	0.1	&	1.107	\\
12-192149	&	Y	&	S0	&	R	&	0.593	&	0.589	&	2.3	&	0.791	&	0.689	&	45.6	&	52.0	&	10.0	&	10.18	&	1.888	&	0.0275	&	1.2	&	24	&	0.099	&	-0.434	&	-0.14	&	0.0	&	0.0	\\
12-110756	&	Y	&	S0	&	R	&	0.555	&	0.595	&	2.2	&	0.961	&	0.728	&	136.1	&	162.5	&	23.5	&	9.92	&	1.74	&	0.0285	&	0.5	&	25	&	-0.321	&	0.0	&	-0.022	&	0.0	&	0.0	\\
12-129601	&	Y	&	S0	&	R	&	0.649	&	0.467	&	2.3	&	0.42	&	0.144	&	24.2	&	6.0	&	4.0	&	11.01	&	2.217	&	0.0304	&	5.2	&	26	&	0.425	&	0.0	&	0.17	&	0.0	&	0.0	\\
12-129434	&	Y	&	S0	&	R	&	0.679	&	0.483	&	2.4	&	0.514	&	0.289	&	148.5	&	144.5	&	13.5	&	9.51	&	1.9	&	0.0267	&	0.9	&	27	&	-0.487	&	0.0	&	-0.812	&	0.0	&	0.0	\\
12-192116	&	Y	&	I	&	R	&	0.879	&	0.656	&	2.3	&	0.368	&	0.233	&	101.5	&	172.0	&	9.0	&	9.4	&	1.754	&	0.0261	&	2.5	&	28	&	0.788	&	-0.494	&	0.037	&	0.2	&	0.392	\\
\bottomrule
\label{tab:MaNGA_table}
\end{tabular}}
\end{sidewaystable*}

\begin{sidewaystable*}
\centering
\caption{Table containing the characteristics for all galaxies in our group catalogue that are not in the MaNGA sample. For each galaxy, columns (1) gives its enclosing group number and columns (2) and (3) give the RA and Dec from the combined NSA+SDSS catalogue Columns (4) - (8) give $\log(\Sigma_3)$, $\log(\Sigma_3^{\rm{rel}})$, $\log(\Sigma_{10})$, $D_{\rm{cen}}$ and $D_{\rm{cen}}^{\rm{norm}}$ which are all measured in this work. Columns (9) and (10) list the redshift and redshift error respectively from the combined catalogue, and Columns (11) and (12) give the absolute magnitude and flux error respectively. Column (13) gives the stellar mass in log units calculated using the relation shown in \autoref{fig:r_band_stellar_mass}. Column (14) gives the ellipticity $\epsilon$ taken from the combined NSA+SDSS catalogue and Column (15) gives the angular momentum classifcation. This classification is S for galaxies which have been visually classified as SRs and satisfy the criteria for SRs, and F otherwise. Finally, Column (16) gives the object name from the HyperLEDA catalogue \protect{\citep{paturel2003hyperleda}} if there is a match. Only the first 20 rows are shown.}
\resizebox{1.000\textwidth}{!}{%
\begin{tabular}{@{}cccccccccccccccc@{}}
\toprule
\begin{tabular}[c]{@{}c@{}}Group\\Number\\(1)\end{tabular} &
\begin{tabular}[c]{@{}c@{}}RA\\(\degree)\\(2)\end{tabular} &
\begin{tabular}[c]{@{}c@{}}DEC\\(\degree)\\(3)\end{tabular} &
\begin{tabular}[c]{@{}c@{}}$\Sigma_3$\\(Mpc$^{-2}$)\\(4)\end{tabular} &
\begin{tabular}[c]{@{}c@{}}$\Sigma_3^{\rm{rel}}$\\(Mpc$^{-2}$)\\(5)\end{tabular} &
\begin{tabular}[c]{@{}c@{}}$\Sigma_{10}$\\(Mpc$^{-2}$)\\(6)\end{tabular} &
\begin{tabular}[c]{@{}c@{}}$D_{\rm{cen}}$\\(Mpc)\\(7)\end{tabular} &
\begin{tabular}[c]{@{}c@{}}$D_{\rm{cen}}^{\rm{norm}}$\\\\(8)\end{tabular} &
\begin{tabular}[c]{@{}c@{}}Redshift\\$z$\\(9)\end{tabular} &
\begin{tabular}[c]{@{}c@{}}Redshift error\\$\Delta z$\\(10)\end{tabular} &
\begin{tabular}[c]{@{}c@{}}Absolute\\Magnitude $M_r$\\(11)\end{tabular} &
\begin{tabular}[c]{@{}c@{}}Flux error\\$\Delta M_r$\\(12)\end{tabular} &
\begin{tabular}[c]{@{}c@{}}$\log(\rm{M})$\\\\(13)\end{tabular} &
\begin{tabular}[c]{@{}c@{}}Ellipticity\\$\epsilon$\\(14)\end{tabular} &
\begin{tabular}[c]{@{}c@{}}Angular\\Momentum\\(15)\end{tabular} &
\begin{tabular}[c]{@{}c@{}}Object\\Name\\(16)\end{tabular} \\ \midrule
6	&	231.05952	&	42.39746	&	0.311	&	0.0	&	-0.054	&	0.4	&	1.061	&	0.0186	&	0.0	&	-19.86	&	0.2	&	10.64	&	0.61	&	F	&	PGC054995	\\
15	&	232.16464	&	42.43981	&	0.136	&	-0.111	&	0.289	&	0.0	&	0.0	&	0.1727	&	0.1098	&	-18.59	&	0.916	&	10.11	&	0.18	&	F	&	-	\\
15	&	232.21886	&	42.42426	&	0.247	&	0.0	&	0.304	&	0.3	&	1.214	&	0.0751	&	0.0	&	-19.83	&	0.2	&	10.63	&	0.26	&	F	&	SDSSJ152852.52+422527.3	\\
21	&	205.47557	&	27.0626	&	0.595	&	-0.152	&	-0.04	&	0.0	&	0.0	&	0.0287	&	0.0	&	-18.52	&	0.2	&	10.08	&	0.84	&	F	&	PGC1796683	\\
21	&	205.54325	&	26.8366	&	0.42	&	-0.327	&	-0.045	&	0.5	&	1.271	&	0.029	&	0.0	&	-18.82	&	0.2	&	10.21	&	0.2	&	F	&	PGC1789796	\\
25	&	203.93349	&	26.11623	&	0.712	&	0.0	&	0.483	&	0.0	&	0.0	&	0.0677	&	0.0	&	-21.12	&	0.2	&	11.17	&	0.42	&	F	&	PGC1763246	\\
25	&	203.96621	&	26.11137	&	0.454	&	-0.258	&	0.561	&	0.2	&	0.441	&	0.0641	&	0.0225	&	-18.6	&	0.034	&	10.11	&	0.27	&	F	&	-	\\
25	&	203.84106	&	26.10623	&	0.509	&	-0.203	&	0.299	&	0.4	&	1.239	&	0.075	&	0.0101	&	-20.68	&	0.223	&	10.98	&	0.35	&	F	&	PGC1762840	\\
27	&	205.26734	&	27.69384	&	0.281	&	-0.022	&	-0.118	&	0.0	&	0.0	&	0.0286	&	0.0	&	-19.68	&	0.2	&	10.57	&	0.51	&	F	&	PGC1815439	\\
29	&	203.69127	&	25.44989	&	0.156	&	-0.377	&	-0.138	&	0.2	&	0.535	&	0.0264	&	0.0	&	-18.25	&	0.2	&	9.97	&	0.53	&	F	&	PGC1736968	\\
29	&	203.77524	&	25.71026	&	0.533	&	0.0	&	-0.102	&	0.4	&	1.116	&	0.0485	&	0.0296	&	-18.23	&	0.008	&	9.96	&	0.43	&	F	&	-	\\
34	&	204.60071	&	26.77924	&	1.282	&	0.0	&	-0.015	&	0.0	&	0.0	&	0.0291	&	0.0	&	-19.17	&	0.2	&	10.35	&	0.41	&	F	&	PGC048203	\\
34	&	204.60445	&	26.74248	&	1.051	&	-0.231	&	-0.022	&	0.1	&	0.173	&	0.0285	&	0.0	&	-22.07	&	0.2	&	11.57	&	0.18	&	S	&	IC4314	\\
34	&	204.53374	&	26.86278	&	0.788	&	-0.494	&	0.041	&	0.2	&	0.5	&	0.0287	&	0.0	&	-18.78	&	0.2	&	10.19	&	0.4	&	F	&	PGC1790602	\\
37	&	317.04129	&	11.3196	&	0.394	&	-0.071	&	0.261	&	0.0	&	0.0	&	0.0522	&	0.0364	&	-18.57	&	1.816	&	10.1	&	0.45	&	F	&	-	\\
37	&	317.03331	&	11.3562	&	0.436	&	-0.029	&	0.164	&	0.2	&	0.244	&	0.0713	&	0.0	&	-20.96	&	0.2	&	11.1	&	0.49	&	F	&	PGC1392380	\\
37	&	317.15099	&	11.38284	&	0.394	&	-0.071	&	0.314	&	0.6	&	0.823	&	0.0755	&	0.0196	&	-19.02	&	0.048	&	10.29	&	0.65	&	F	&	-	\\
37	&	316.99296	&	11.18472	&	0.396	&	-0.069	&	0.313	&	0.7	&	0.931	&	0.0442	&	0.0342	&	-19.93	&	0.006	&	10.67	&	0.7	&	F	&	-	\\
37	&	317.06798	&	11.16976	&	0.312	&	-0.153	&	0.383	&	0.7	&	0.989	&	0.1268	&	0.0563	&	-19.3	&	0.011	&	10.4	&	0.93	&	F	&	-	\\
37	&	317.16755	&	11.412	&	0.21	&	-0.255	&	0.207	&	0.8	&	1.016	&	0.0714	&	0.0	&	-20.62	&	0.2	&	10.96	&	0.8	&	F	&	PGC1393205	\\
\bottomrule
\label{tab:Not_MaNGA_table}
\end{tabular}}
\end{sidewaystable*}


\end{document}